   \documentclass[structabstract]{aa}
\usepackage{graphicx}
\usepackage{txfonts}
\usepackage{natbib}
\usepackage{ulem}
\usepackage{textcomp}
 
\usepackage{bm}
\usepackage{multirow}

\bibpunct[]{(}{)}{;}{a}{}{,}

\begin{document}

   \title{Laboratory rotational spectroscopy of isotopic acetone, CH$_3^{13}$C(O)CH$_3$ and 
          $^{13}$CH$_3$C(O)CH$_3$, and astronomical search in \object{Sagittarius~B2(N2)}\thanks{
The experimental line lists are available at the CDS via anonymous ftp
to cdsarc.u-strasbg.fr (130.79.128.5) or via
\protect\url{http://cdsweb.u-strasbg.fr/cgi-bin/qcat?J/A+A/}}}

   \author{Matthias H. Ordu\inst{1}
           \and
           Oliver Zingsheim\inst{1}
           \and
           Arnaud Belloche\inst{2}
           \and
           Frank Lewen\inst{1}
           \and
           Robin T. Garrod\inst{3}
           \and
           Karl M. Menten\inst{2}
           \and
           Stephan Schlemmer\inst{1}
           \and
           Holger S.~P. M{\"u}ller\inst{1}
                   }

   \institute{I.~Physikalisches Institut, Universit{\"a}t zu K{\"o}ln,
              Z{\"u}lpicher Str. 77, 50937 K{\"o}ln, Germany\\
              \email{hspm@ph1.uni-koeln.de}
         \and
              Max-Planck-Institut f\"ur Radioastronomie, Auf dem H\"ugel 69, 
              53121 Bonn, Germany
         \and
              Departments of Chemistry and Astronomy, University of Virginia, Charlottesville, 
              VA 22904, USA
              }

   \date{Received 14 May 2019 / Accepted 13 June 2019}

\abstract
{
Spectra of minor isotopic species of molecules that are abundant in space may also be detectable.
Their respective isotopic ratios may provide clues about the formation of these molecules.
Emission lines of acetone in the hot molecular core Sagittarius~B2(N2) are strong enough to warrant 
a search for its singly substituted $^{13}$C isotopologs.
}
{
We want to study the rotational spectra of CH$_3^{13}$C(O)CH$_3$ and
$^{13}$CH$_3$C(O)CH$_3$ and search for them in Sagittarius~B2(N2).
}
{
We investigated the laboratory rotational spectrum of isotopically enriched 
CH$_3^{13}$C(O)CH$_3$ between 40~GHz and 910~GHz and of acetone between 
36~GHz and 910~GHz in order to study $^{13}$CH$_3$C(O)CH$_3$ in natural isotopic composition. 
In addition, we searched for emission lines produced by these species in a molecular 
line survey of Sagittarius~B2(N) carried out with the Atacama Large 
Millimeter/submillimeter Array (ALMA).
Discrepancies between predictions of the main isotopic species and the 
ALMA spectrum prompted us to revisit the rotational spectrum of this isotopolog.
}
{
We assigned 9711 new transitions of CH$_3^{13}$C(O)CH$_3$ and 63 new transitions of $^{13}$CH$_3$C(O)CH$_3$ in the 
laboratory spectra. More than 1000 additional lines were assigned
for the main isotopic species. We modeled the ground state data of all three 
isotopologs satisfactorily with the ERHAM program. We find that models of the torsionally 
excited states $\varv _{12} = 1$ and $\varv _{17} = 1$ of CH$_3$C(O)CH$_3$ 
improve only marginally.
No transition of CH$_3^{13}$C(O)CH$_3$ is clearly detected toward the hot
molecular core Sgr~B2(N2). However, we report a tentative detection
of $^{13}$CH$_3$C(O)CH$_3$ with a $^{12}$C/$^{13}$C isotopic ratio of 27 
that is consistent with the ratio previously measured for alcohols in this source. 
Several dozens of transitions of both torsional states of the main 
isotopolog are detected as well.
}
{
Our predictions of CH$_3^{13}$C(O)CH$_3$ and CH$_3$C(O)CH$_3$ are reliable 
into the terahertz region. The spectrum of $^{13}$CH$_3$C(O)CH$_3$ should be 
revisited in the laboratory with an enriched sample.
The torsionally excited states $\varv _{12} = 1$ and $\varv _{17} = 1$
of CH$_3$C(O)CH$_3$ were not reproduced satisfactorily in our models.
Nevertheless, transitions pertaining to both states could be identified
unambiguously in Sagittarius B2(N2).
}

\titlerunning{Spectroscopy of isotopic acetone in the
laboratory and in Sgr~B2(N)}

\maketitle
\hyphenation{For-schungs-ge-mein-schaft}

%

\section{Introduction}
\label{intro}

Among the more than 200 different molecules detected in the interstellar medium 
(ISM) or in circumstellar envelopes, there are several organic molecules 
containing three or more heavy atoms\footnote{See, e.\,g., the Molecules in Space 
page, https://cdms.astro.uni-koeln.de/classic/molecules, of the Cologne Database 
for Molecular Spectroscopy, CDMS}. Several of these are so abundant that 
minor isotopic species were detected. Recent detections of isotopologs 
containing $^{13}$C include ethyl cyanide \citep{13C-EtCN_2007}, 
vinyl cyanide \citep{13C-VyCN_2008}, methyl formate \citep{13C-MeFo_2009}, 
dimethyl ether \citep{13C-DME_2013}, ethanol \citep{13C-EtOH_det_2016,13C-EtOH_rot_2012}, 
glycoladehyde \citep{13C-Glycald_det_2016,13C-Glycald_rot_2013}, formic acid 
\citep{PILS_13C_D_div_2018,HCOOH_rot_2008}, ketene 
\citep{PILS_13C_D_div_2018,ketene_rot_2003}, and acetaldehyde 
\citep{PILS_13C_D_div_2018,13C-MeCHO_rot_2015}.
Even the doubly $^{13}$C substituted isotopologs were detected in the case of 
methyl cyanide \citep{2x13C-MeCN_det_2016,CH3CN_rot_2009} and ethyl cyanide 
\citep{2x13C-EtCN_2016}. These detections relied heavily on available laboratory 
spectra, which  were frequently published concomitantly. In the cases where 
the necessary laboratory spectra were published earlier, these are given as 
second, older reference, in the list above.

Acetone, or propanone, is one of the most complex molecules found in the ISM. 
Its first detection toward Sagittarius (Sgr) B2(N) was reported 30 years ago 
by \citet{acetone_det_1987}. The derived column densities were considerably 
higher than predicted by astrochemical models at that time, with the consequence 
that the detection was widely disputed. The detection was confirmed 15 years 
later \citep{acetone_conf_2002}. Rotational transitions of acetone, including 
some in the first excited torsional mode, were also detected in the hot core 
associated with Orion~KL \citep{acetone_Orion-KL_2005}. The molecule was 
detected very recently toward the low mass protostar IRAS 16293$-$2422B 
\citep{acetone_16293_2017}.

Some of the co-authors of this paper used the IRAM 30~m telescope to carry out 
an unbiased molecular line survey toward Sgr~B2(N) and Sgr~B2(M) at 3~mm 
wavelength with additional measurements at shorter wavelengths 
\citep{SgrB2N_30m_2013}. The survey led to the detection of aminoacetonitrile 
\citep{det_AAN}, ethyl formate, and \textit{n}-propyl cyanide \citep{det-PrCN_EtFo}, 
among other interesting results. Acetone was detected toward both sources 
\citep{SgrB2N_30m_2013}. The emission lines toward Sgr~B2(N) were strong enough 
that the $^{13}$C isotopologs may be identifiable, if not in our IRAM 30~m data, 
then in the 3~mm survey carried out with the Atacama Large Millimeter/submillimeter 
Array (ALMA) in its Cycles~0 and 1 \citep{2x13C-MeCN_det_2016}. 
The survey is called EMoCA, which stands for Exploring Molecular Complexity with ALMA. 
The sensitivity of these data not only permitted the observation of 
\textit{n}-propyl cyanide in its ground vibrational state but also in four 
vibrationally excited states for each of its two conformers \citep{vib_n-PrCN_2016}. 
In addition, detection of \textit{iso}-propyl cyanide as the first branched 
alkyl compound in the interstellar medium was reported \citep{det_i-PrCN_2014}.

The rotational spectrum of the main isotopolog of acetone in its ground vibrational 
state was studied for the first time by \cite{acetone_first_obs}, and until now, 
it was examined fairly extensively (see \citealt{acetone-12C_groner_2002} 
and references therein). In addition, rotational spectra in the low-lying 
torsional states $\varv_{12} = 1$ and $\varv_{17} = 1$ were characterized 
by \citet{acetone_v12eq1_2006,acetone_v17eq1_2008} and very
recently by \citet{acetone-v17_2019}. During the research for this paper, 
we learned about the recent work from \cite{acetone-12C_2016}, but their results 
could not be considered in the present study because  no measured or 
predicted line frequencies are available.

In contrast, no information was available on the singly $^{13}$C substituted 
isotopologs of acetone until \citet{mono-13C-acetone} reported on a 
Fourier transform microwave (FTMW) spectroscopic investigation of a 
jet-cooled sample of acetone in its natural isotopic composition. 
The data are sufficient for a basic characterization of both isotopomers,
but they are not sufficient for an extrapolation into the 3~mm wavelength range. 
Therefore, this study embarks on the millimeter- and submillimeter-wavelength 
spectra of the singly $^{13}$C substituted isotopologs of acetone.

We focused our investigation on the symmetric isotopolog, CH$_3^{13}$C(O)CH$_3,$ 
because an isotopically enriched sample was commercially available.
The analysis proved to be considerably more challenging than for
dimethyl ether \citep{DME_rot_2009} or for its isotopologs with one and two 
$^{13}$C \citep{13C-DME_2013}. Additionally, since we faced difficulties in obtaining 
an enriched sample of the asymmetric isotopolog, we tried to make assignments 
in a sample of natural isotopic composition for $^{13}$CH$_3$C(O)CH$_3$, 
albeit with limited success.
Difficulties with modeling the main isotopic species of acetone in its ground 
vibrational state in the ALMA data and the need for predictions in the torsionally 
excited states prompted further investigation of this isotopolog, which used the
experience gained in the analysis of the symmetric $^{13}$C isotopolog.

\section{Experimental details}
\label{exptl}

\begin{figure}[t]
\centering
\includegraphics[width=\linewidth]{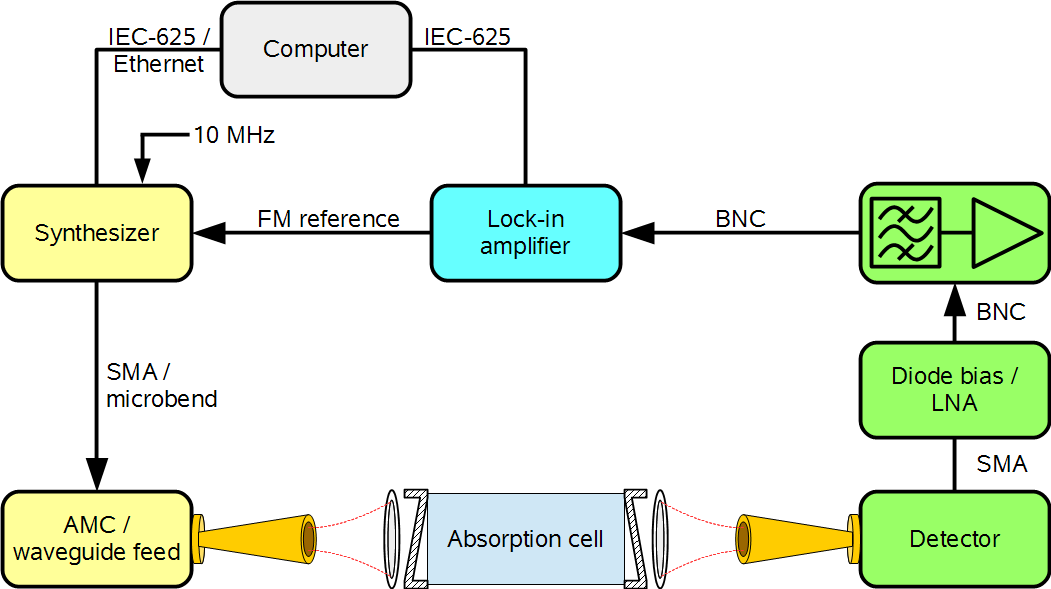}
\caption{Schematic of the general experimental setup.
A single-pass setup is shown for the sake 
of clarity, although several measurements were performed with a double-pass setup,
where the frequency source and detector are located on the
same side of the cell.}
\label{schematic_exp_setup}
\end{figure}

We performed measurements with the MIllimeter-wave Double-pass
Absorption Spectrometer for COmplex INterstellar Species 
(MIDAS-COINS; setup 1) and with a submillimeter spectrometer 
(setup 2) at room temperature. A schematic of the general experimental 
setup can be found in Fig.~\ref{schematic_exp_setup}. 
The experimental setup primarily consists  of the following three parts: 
an oscillator generating a polarized electromagnetic wave 
(frequency source), an absorption cell, and a detector.
The selection of the frequency source and the detector 
strongly depend on the desired frequency region. 
An overview of the devices used for measuring the spectra of
CH$_3^{13}$C(O)CH$_3$, the main focus of this work, is listed in 
Table~\ref{exp_setups}. 
The sources of Setup~1 were commercially available 
synthesizers with different amplifier-multiplier chains (AMCs)
for the frequency bands around 100 and 200~GHz. 
GaAs Schottky diodes were used as detectors.
The diodes were supplied with a negative bias current by in-house fabricated bias boxes,
which are equipped with ultra-low-noise preamplifiers. 
The waveguide, containing the Schottky diode that was used for the
100~GHz band, had a tunable back-short
for optimizing the detector response (impedance matching).
The back-short was retuned about 
every gigahertz in order to maintain optimum signal recovery.  
Setup~1 is also suitable for higher frequencies, 
but it is outperformed by bolometric detection techniques. 
Therefore, we
used a cryogenic InSb bolometer 
with a built-in bias box (Setup~2) for the frequency ranges from 270 to 910~GH . 
Additionally, we used a superlattice multiplier from a collaboration with D.~G.~Paveliev 
\citep{superlattice_multiplier}, a backward-wave oscillator, described 
in detail by \citet{BWO}, and a commercially available modular AMC system 
as frequency sources in the submillimeter region. The experimental setup MIDAS-COINS can
have a total absorption path length of up to 44~m.
This long path is particularly useful for studying weak absorption features 
of complex molecules.
In all setups, the absorption cell consisted of borosilicate tubes
with a 10 cm diameter. The cells were vacuum sealed with PTFE o-rings
and windows that were tilted by 10$^{\circ}$ to reduce standing waves. 
The double-pass setup was realized with a roof-top mirror which
rotates the polarization of the electromagnetic wave
such that a beam splitter can separate the
in- and outcoming beams.

The electromagnetic wave was frequency modulated (FM; $f\leq 50$~kHz) 
with a $2f$ demodulation which causes Doppler limited absorption signals 
to appear close to a second derivative of a Gaussian. 
The modulation amplitude was $\leq 0.8~\times HWHM$. 
A 10~MHz rubidium atomic clock was used as a frequency standard.
The accuracy of transition frequencies is mainly limited by standing waves, 
asymmetric line shapes also originating from nearby lines in a dense spectrum,  
and the linewidth which is dominated by Doppler-broadening. 
The uncertainty of a transition frequency $\Delta f$ in a dense spectrum, 
such as that of acetone, usually increases for increasing frequencies 
since the Doppler width is proportional to the frequency of the 
electromagnetic wave. 
The uncertainties of transition frequencies were chosen to be
$\Delta f= HWHM / 6.6$;
the value of 6.6 was determined as the average number of
frequency steps needed to record one $HWHM$ properly. 
A satisfactory signal-to-noise ratio (SNR) was typically
achieved with 20~ms integration time 
per frequency step. To prevent systematic shifts in the
frequency domain, we always averaged an upward scan and a
downward scan (measuring a frequency window with increasing
and decreasing frequencies, respectively).


We studied CH$_3^{13}$C(O)CH$_3$ using an isotopically enriched 
sample from Sigma-Aldrich with 99\% $^{13}$C on the 
carbonyl C and a chemical purity of 99\%. We recorded its spectrum 
at room temperature under static conditions, at a constant
gas pressure of 5~\textmu bar. 
We measured the main isotopolog of acetone, CH$_3$C(O)CH$_3$, 
and $^{13}$CH$_3$C(O)CH$_3$ with a pure sample of acetone in 
natural abundance. 
Because the relative natural abundance ratio is 87:2:1 for 
CH$_3$C(O)CH$_3$:$^{13}$CH$_3$C(O)CH$_3$:CH$_3^{13}$C(O)CH$_3$, 
we measured $^{13}$CH$_3$C(O)CH in natural abundance in a 
constant gas flow at comparably high pressures up to 14~\textmu bar. 
In order to increase the SNR further, the modulation amplitude 
was increased up to $(1.0-1.5) \times HWHM$, and the total integration 
time per point was increased to 1.8~s.

\begin{table*}
\begin{center}

  \caption{Measurement setups for CH$_3^{13}$C(O)CH$_3$.}
  \label{exp_setups}
{\footnotesize
  \begin{tabular}{cr@{}llcl}
 \hline\hline

Setup & \multicolumn{2}{c}{Frequency Range} & RF Source &
Absorption Path Length & Detector  \\
     \hline
1a & 40$-$&~48 GHz & 70 GHz synthesizer & 2 $\times$ 7 m & Schottky diode \\                         
\hline
1b & 75$-$&130 GHz & 43.5 GHz synthesizer & 2 $\times$ 7 m & Schottky diode \\
   &    &        & + AMC ($\times 3$)              & & + tunable backshort \\
   \hline
1c &200$-$&232 GHz & 43.5 GHz synthesizer & 2 $\times$ 7 m & Schottky diode \\                                                  
   &    &        & + AMC ($\times 16$)              & &  \\
      \hline
2a &270$-$&304 GHz & Source 1b  & 3.4 m & 4 K InSb bolometer \\                                                  
   &    &        & + superlattice multiplier               & &  \\
    \hline    
2b &335$-$&365 GHz & Backward-wave oscillator  & 3.4 m & 4 K InSb bolometer \\                                                  
    \hline   
2c &629$-$&710 GHz & Microwave synthesizer  & 3.4 m & 4 K InSb bolometer \\                                                  
\cline{1-3}
2d &840$-$&910 GHz & + modular AMC system   &   \\
    \hline   
  \end{tabular}\\[2pt]
}
\end{center}

\end{table*}

\section{Rotational spectroscopy of acetone}
\subsection{Spectroscopic properties of acetone}
\label{spec-props}

Acetone is a rather asymmetric oblate rotor with
$\kappa = (2B - A - C)/(A - C) = +0.3720$, which is
quite far from the symmetric limit of $+1$.
The symmetry of the rigid molecule is $C_{\rm 2v}$, 
and the dipole moment component of $2.93\pm0.03$~D
\citep{acetone_dipolemoment} is along 
the $b$-inertial axis of the molecule.
The selection rules are $\Delta J=0, \pm 1$, 
$\Delta K_a \equiv 1$\,mod(2), and $\Delta K_c \equiv 1$\,mod(2).

The effective barrier to internal rotation of the two equivalent methyl rotors is moderately 
low, 251~cm$^{-1}$ , 361~K, or 3.00~kJ/mol \citep{acetone_barrier_2000}, such that most 
rotational transitions are split into four lines. 
These components are labeled by the
quantum numbers of the internal rotation Hamiltonian of the two methyl groups in the absence of external electromagnetic fields, $\sigma_1$ and $\sigma_2$, respectively. 
Rotational transitions occur only within a given internal rotation component.
Relative intensities for allowed \textit{b}-type transitions can be derived by group-theory 
considerations and are listed in Table~\ref{Table_Int-acetone}.

The two equivalent methyl rotors lead to two torsional
modes whose fundamental frequencies
are quite low in acetone; for the symmetric torsional mode $\varv _{12} = 1$
is at 77.8~cm$^{-1}$ or 112~K 
\citep{acetone_v12_1992} while for the asymmetric torsional mode
$\varv _{17} = 1$ is at 124.6~cm$^{-1}$ or 179~K \citep{acetone_FIR_1987}.
Rotational transitions pertaining to these states are not much weaker 
than the ground state transitions at room temperature. 
This leads to a very rich rotational spectrum within a given vibrational 
state and a plethora of vibrational states that are stronger than 
the $^{13}$C isotopologs in natural isotopic composition.

Substitution of the carbonyl C atom by $^{13}$C, resulting in 
CH$_3^{13}$C(O)CH$_3$, retains the symmetry of the molecule. 
Moreover, the substitution has very small effects on the lower 
order spectroscopic parameters because the atom is close to the 
center of mass. In contrast, substitution of one methyl C atom 
by $^{13}$C, resulting in $^{13}$CH$_3$C(O)CH$_3$, lowers the symmetry 
and leads to changes even in the lower order spectroscopic parameters.
This includes the internal rotation parameters associated with the two 
different rotors. However, the changes are small and may be neglected 
if the values are not determined with high accuracy. The number of 
internal rotation components is now five.


\begin{table}
\begin{center}
\caption{Relative intensities resulting from internal rotation of both methyl groups in acetone.}
  \label{Table_Int-acetone}
{\footnotesize
  \begin{tabular}{r@{}r@{}lr@{}r@{}lccc}
 \hline\hline
\multicolumn{6}{c}{Torsional substate} & \multicolumn{2}{c}{CH$_3$C(O)CH$_3$}  &
\multicolumn{1}{c}{$^{13}$CH$_3$C(O)CH$_3$} \\
 & & & & & & \multicolumn{2}{c}{CH$_3^{13}$C(O)CH$_3$}  &   \\     
  ($\sigma_1$,& ~$\sigma_2$)&\tablefootmark{a} &
  ($\sigma_1$,& ~$\sigma_2$)&\tablefootmark{b} &
$ee \leftrightarrow oo$\tablefootmark{c} &
$eo \leftrightarrow oe$\tablefootmark{c} \\
     \hline
     (~0,& 0)          && (~0,& 0) && 3 & 5 & 2  \\
     ($\pm$1,& 0)      && (~1,& 0) && \multirow{2}{*}{8} & \multirow{2}{*}{8} & 2  \\
     (0,& $\pm$1)      && (~0,& 1) &&   &   & 2  \\
     ($\pm$1,& $\pm$1) && (~1,& 1) && 1 & 3 & 1  \\
     ($\pm$1,& $\mp$1) && (~1,& 2) && 2 & 2 & 1  \\                       

    \hline    
  \end{tabular}\\
}
\end{center}
   \tablefoot{
 \tablefoottext{a}{Quantum numbers of the
internal rotation Hamiltonian of the two methyl groups in the absence of external electromagnetic fields,
 $\sigma_1$ and $\sigma_2$, respectively.}
 \tablefoottext{b}{ERHAM notation.}
 \tablefoottext{c}{Selection rules for transitions from energy
 levels with even (e) and odd (o) quantum numbers of $K_a$ and
 $K_c$ ($K_a'~K_c' \leftrightarrow K_a''~K_c''$).} 
 }
\end{table}

\subsection{Spectroscopic analysis}

\label{lab-obs}

We used the software package Assignment and Analysis of Broadband Spectra 
(AABS) developed by Z. Kisiel \citep{AABS1,AABS2} for analyzing the 
experimental spectra. We employed an extended version of the ``Effective 
Rotational HAMiltonian program for molecules with up to two periodic 
large-amplitude motions'' (ERHAM) from P. Groner 
\citep{ERHAM,acetone-12C_groner_2002} for modeling because we needed 
to fit large quantum numbers to experimental uncertainty efficiently. 
The extension allowed us to process more transitions and parameters 
from the input file and the Intel Fortran compiler partially 
optimized the program for parallel calculation.

The ERHAM code uses an effective rotational Hamiltonian for fitting one or 
two periodic large-amplitude motions (LAMs). It does not solve the 
internal-rotation Hamiltonian itself, in particular no barriers to internal 
rotation are calculated. The program uses the periodicity of eigenfunctions 
to expand them as a Fourier series. Its coefficient functions are dependent 
on the internal rotation variable. Furthermore, all matrix elements of the 
full Hamiltonian can be expanded as a Fourier series. The Hamiltonian $H$ is of 
the form $H=T+V$, with $T$ describing the kinetic energy and $V$ describing 
the potential energy. Expressing overall rotation and internal rotation 
of two LAMs in a general form, $T$ is 
\begin{equation}
T=\frac{1}{2}\left( \bm{\tilde{\omega}} \bm{\tilde{\dot{\tau}}} \right)  
\begin{pmatrix}
\bm{I}_\omega & \bm{I}_{\omega\tau}  \\
\bm{\tilde{I}}_{\omega\tau} & \bm{I}_\tau
\end{pmatrix}
\begin{pmatrix}
\bm{\omega}   \\
\bm{\dot{\tau}}
\end{pmatrix}
\end{equation} 
with $\bm{\omega}$ the overall angular velocity vector 
$(3 \times 1$-matrix), $\bm{\dot{\tau}}$ the velocities of both 
internal motions $(2 \times 1)$, $\bm{I}_\omega$ the moment of 
inertia of overall rotation $(3 \times 3)$, $\bm{I}_\tau$ 
containing the moments of inertia of the internal motions 
on its diagonal $(2 \times 2)$, and  $\bm{I}_{\omega\tau}$ 
containing the cross terms $(3 \times 2)$. 


The overall fitting procedure was as follows: after a round of new assignments, 
we carried out new fits with ERHAM under systematic variations of parameters 
to reach a weighted standard deviation close to 1.0, which means that the 
experimental data were fit within their uncertainties on average. 
Too many optimistic uncertainties, wrong assignments, or an incomplete or wrong 
model may cause values much larger than 1.0, whereas too many conservative 
uncertainties may lead to values smaller than 1.0. Transitions with large 
residuals after the fit (more than about ten times the uncertainties) were 
weighted out at this stage because of potential misassignments. 
Consequently, for each applicable combination of localized states of the 
first and second motions, labeled $(q,q')$, we tested possible missing 
parameters by including all parameters on the same order of $n=K+P+R$ 
to the fit; $K$ defines the order of $\textbf{P}^K$, $P$ of 
$\textbf{P}_z^P$ , and $R$ of $(\textbf{P}_+^R + \textbf{P}_-^R)$. 
Next, sets of parameters with higher orders of $n$,
as well as sets of parameters of ascending order of $(q+q')$, were added 
as long as the standard deviation improved by at least 10\%. 
Subsequently, we searched for parameters which could be omitted from the fit 
without a significant increase in the weighted standard deviation of the fit. 
These were mostly parameters with large uncertainties compared to 
the magnitudes of their values. A new round of assignments was started 
after completion of this procedure until no new lines could be assigned 
in any of the spectral recordings.

At an advanced stage of the assignment process, we noticed that 
parameters with an odd order of $n$ turned out to be indispensable. 
All parameters not in line with Watson's A-reduction \citep{A-reduction}
nor with the S-reduction \citep{S-reduction_experimental,S-reduction}
are labeled in cylindrical tensor form
with $[B_{KPR}]_{qq'}$ or $[B^-_{KPR}]_{qq'}$ (the superscript $-$ is to
highlight that $\omega = -1$) according to the tunneling parameter input
summarized by \cite{ERHAM_overview} in Appendix A.
Such parameters were used rarely in previous fits carried out with 
the ERHAM program. In addition, they also helped to obtain a greatly 
improved fit of the main isotopic species in its ground vibrational state.

\section{Results}
\subsection{Laboratory spectroscopy}

We studied the ground vibrational states of the symmetric and asymmetric 
isotopologs CH$_3^{13}$C(O)CH$_3$ and $^{13}$CH$_3$C(O)CH$_3$, respectively. 
In addition, we investigated the ground as well as the first two torsionally 
excited states of the main isotopic species.

\subsubsection{The symmetric isotopolog CH$_3^{13}$C(O)CH$_3$
in its ground vibrational state}
\label{13C-sym_v_0}

\citet{mono-13C-acetone} determined 44 transition frequencies of 11 different rotational 
transitions with $J \le 4$ in the region of 10$-$25~GHz. The resulting spectroscopic 
parameters were accurate enough to assign low-$J$ transitions up to 130~GHz. 
We were able to extend the assignments subsequently on the basis of improved
predictions of the rotational 
spectrum to somewhat higher $J$ and $K$ quantum numbers, eventually reaching 
$J=92$ and $K=45$. Inclusion of the tunneling parameters 
$[B^-_{010}]_{10}$ ($=[g_a]_{10}$) and $[B^-_{001}]_{10}$ ($=[g_b]_{10}$) 
in the fit lowered the standard deviation by 17\%. 
Our final line list consists of 5870 lines which were reproduced 
with a weighted standard deviation of 1.29 using 51 spectroscopic 
parameters, see Table~\ref{spectroscopic-parameters-2-13C}.


\begin{table}
\begin{center}

  \caption{Ground state spectroscopic parameters\tablefootmark{a} of CH$_3^{13}$C(O)CH$_3$.}
  \label{spectroscopic-parameters-2-13C}
{\footnotesize
  \begin{tabular}{lr@{}lr@{}l}
 \hline\hline

 Parameter & \multicolumn{2}{c}{Lovas \& Groner} & \multicolumn{2}{c}{This work} \\
              & \multicolumn{2}{c}{(2006)}          &  \\        
     \hline
     \\          

                     $\rho$    &       0&.062074(27)    &       0&.0614858~(41)   \\
 $\beta~/^{\circ}$    &      25&.8224(33)      &      25&.7140~(31)      \\
                             $A$ ~/MHz    &   10164&.00791(76)     &   10164&.005782~(145)   \\
                             $B$ ~/MHz    &    8516&.08462(99)     &    8516&.083092~(104)   \\
                             $C$ ~/MHz    &    4910&.23681(74)     &    4910&.235399~(107)   \\
                    $\Delta_{J}$ ~/kHz    &       4&.957(98)       &       4&.852046~(73)    \\
                   $\Delta_{JK}$ ~/kHz    &    $-$3&.08(12)        &    $-$3&.192449~(194)   \\
                    $\Delta_{K}$ ~/kHz    &       9&.829(94)       &       9&.85055~(33)     \\
                    $\delta_{J}$ ~/kHz    &       2&.042(16)       &       2&.0381002~(265)  \\
                    $\delta_{K}$ ~/kHz    &    $-$0&.617(61)       &    $-$0&.619619~(127)   \\
                      $\Phi_{J}$ ~/mHz    &      50&.6046\tablefootmark{b}      &       6&.1005~(146)     \\
                     $\Phi_{JK}$ ~/mHz    &  $-$336&.741\tablefootmark{a}       &   $-$31&.691~(67)       \\
                     $\Phi_{KJ}$ ~/mHz    &       0&.\tablefootmark{a}          &      45&.10~(34)        \\
                      $\Phi_{K}$ ~/mHz    &     423&.395\tablefootmark{a}       &      22&.83~(36)        \\
                      $\phi_{J}$ ~/mHz    &      25&.3760\tablefootmark{b}      &       3&.1559~(71)      \\
                     $\phi_{JK}$ ~/mHz    &   $-$27&.3291\tablefootmark{b}      &      37&.600~(73)       \\
                      $\phi_{K}$ ~/mHz    &  $-$221&.468\tablefootmark{b}       &   $-$75&.839~(195)      \\
                $L_{KKJ}$ ~/\textmu Hz    &        &               &    $-$2&.551~(156)      \\                   
                  $L_{K}$ ~/\textmu Hz    &        &               &       1&.965~(165)      \\
                         $l_{J}$ ~/nHz    &        &               &    $-$6&.98~(54)        \\ 
                        $l_{JK}$ ~/nHz    &        &               &     299&.3~(86)         \\                                  
                        $l_{KJ}$ ~/nHz    &        &               &     481&.~(47)          \\
                         $l_{K}$ ~/nHz    &        &               &     777&.~(91)          \\

                 $\epsilon_{10}$ ~/MHz    &  $-$763&.36(33)        &  $-$763&.9260~(53)      \\
                 $\epsilon_{20}$ ~/MHz    &       0&.766643\tablefootmark{b}    &       0&.76612~(173)    \\
                 $\epsilon_{11}$ ~/MHz    &       1&.049511\tablefootmark{b}    &       1&.1059~(61)      \\
                $\epsilon_{1-1}$ ~/MHz    &       0&.0799732\tablefootmark{b}   &       0&.08853~(176)    \\            
           
              $[A-(B+C)/2]_{10}$ ~/kHz    &      60&.6(20)         &      40&.853~(166)      \\
                $[(B+C)/2]_{10}$ ~/kHz    &   $-$18&.73(33)        &    $-$3&.291~(141)      \\
                $[(B-C)/4]_{10}$ ~/kHz    &    $-$2&.17(16)        &       5&.530~(70)       \\

                $[\Delta_J]_{10}$ ~/Hz    &        &               &      29&.276~(174)      \\
             $[\Delta_{JK}]_{10}$ ~/Hz    &        &               &   $-$88&.48~(45)        \\             
              $[\Delta_{K}]_{10}$ ~/Hz    &        &               &      57&.385~(292)      \\
              $[\delta_{J}]_{10}$ ~/Hz    &        &               &       2&.277~(35)       \\
              $[\delta_{K}]_{10}$ ~/Hz    &        &               &   $-$13&.076~(57)       \\
                     $[d_2]_{10}$ ~/Hz    &        &               &   $-$12&.531~(94)       \\
 
                 $[\Phi_J]_{10}$ ~/mHz    &        &               &    $-$6&.580~(47)       \\ 
              $[\Phi_{JK}]_{10}$ ~/mHz    &        &               &      18&.431~(109)      \\
              $[\Phi_{KJ}]_{10}$ ~/mHz    &        &               &   $-$16&.384~(80)       \\                            
               $[\Phi_{K}]_{10}$ ~/mHz    &        &               &       4&.660~(56)       \\ 
               $[\phi_{J}]_{10}$ ~/mHz    &        &               &    $-$0&.8706~(69)      \\
              $[\phi_{JK}]_{10}$ ~/mHz    &        &               &       3&.5242~(194)     \\
               $[\phi_{K}]_{10}$ ~/mHz    &        &               &    $-$2&.629~(30)       \\
                    $[h_2]_{10}$ ~/mHz    &        &               &       2&.9911~(239)     \\
                    $[h_3]_{10}$ ~/mHz    &        &               &       0&.5718 ~(49)     \\

              $[B^-_{010}]_{10}$ ~/kHz    &        &               &  $-$696&.4~(66)         \\
               $[B^-_{030}]_{10}$ ~/Hz    &        &               &  $-$203&.16~(107)       \\
               $[B^-_{012}]_{10}$ ~/Hz    &        &               &  $-$658&.8~(79)         \\
              $[B^-_{212}]_{10}$ ~/mHz    &        &               &     190&.09~(168)       \\
              $[B^-_{001}]_{10}$ ~/kHz    &        &               &  $-$288&.16~(197)       \\
           
                $[(B-C)/4]_{1-1}$ ~/Hz    &        &               &      45&.42~(212)       \\ 
                                                                                             \\

    \hline

 no. of transitions fit                   &     44 &               &    9715&                \\
 no. of lines fit                         &     44 &               &    5870&                \\
 standard deviation\tablefootmark{c}      &       0&.96            &       1&.29             \\

    \hline    
  \end{tabular}\\[2pt]
}
\end{center}
   \tablefoot{
\tablefoottext{a}{Numbers in parentheses are one standard deviation
in units of the least significant figures.} 
\tablefoottext{b}{Fixed values, taken from the main isotopic species.} 
\tablefoottext{c}{Weighted unitless value for the entire fit.} 
}
\end{table}


\subsubsection{The asymmetric isotopolog $^{13}$CH$_3$C(O)CH$_3$
in its ground vibrational state}
\label{13C-asym}

\begin{figure}[t]
\centering
\includegraphics[width=\linewidth]{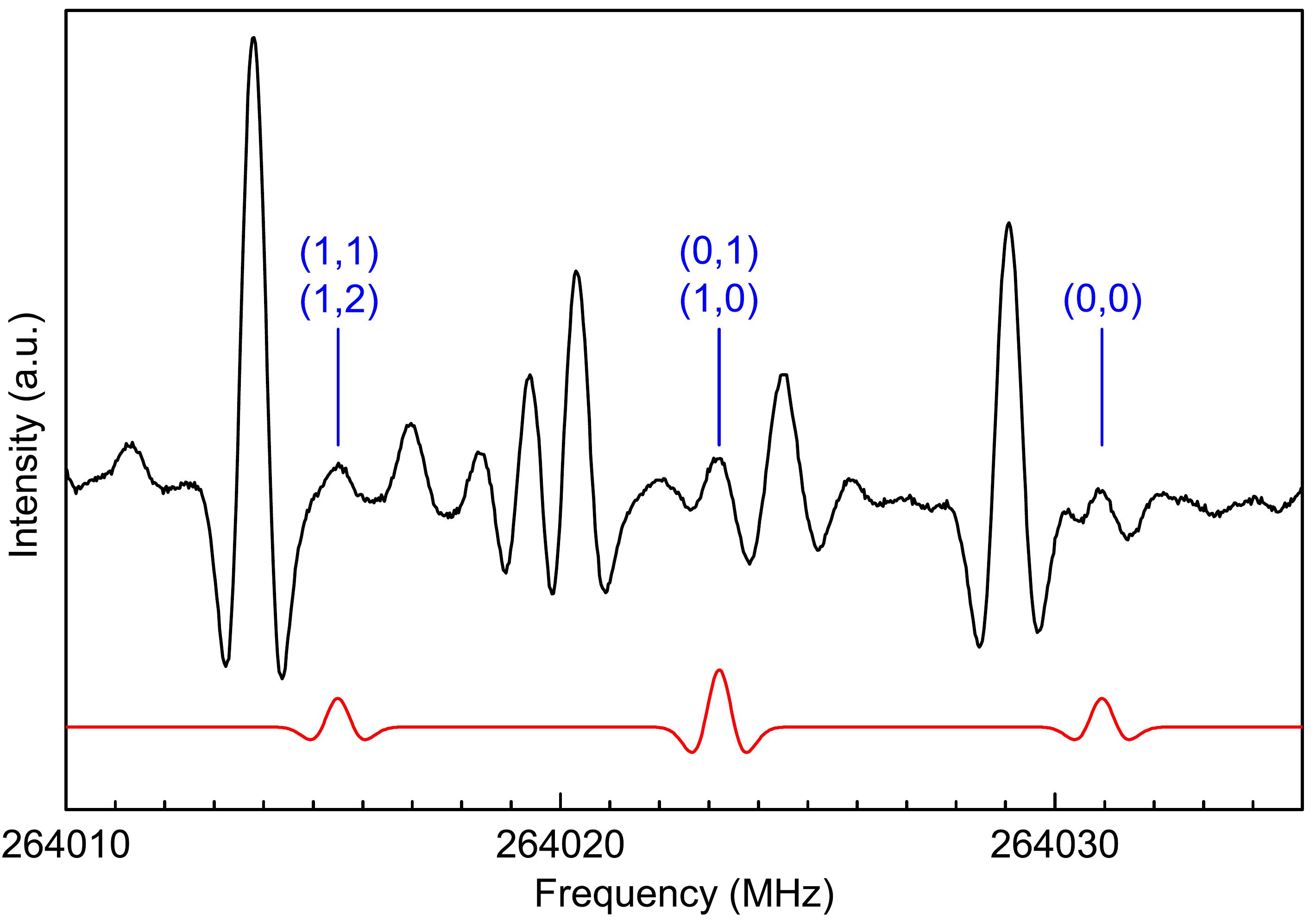}
\caption{Section of the rotational spectrum of acetone in
natural isotopic composition (in black) in the region of
the oblate paired $27_{K_A',27}\leftarrow 26_{K_A'',26}$ transition of
$^{13}$CH$_3$C(O)CH$_3$ with $K_a' = 0,1$ and $K_a" = 1,0$.
The red trace shows the simulation from our final parameters
in Table~\ref{spectroscopic-parameters-1-13C}.
The internal rotation pattern $(1,1)+(1,2):(0,1)+(1,0):(1,1)$ is
particularly easy to identify
because of its simple $(1+1):(2+2):2 = 1:2:1$ intensity ratio and the
torsional substates are given
in the ERHAM notation (in blue), see
Table~\ref{Table_Int-acetone}.
No other line in the shown section can be assigned based on the data from our
present study. Thus, the dense spectrum also demonstrates
the challenge of assigning $^{13}$CH$_3$C(O)CH$_3$ lines in natural abundance.}
\label{spectra_13C}
\end{figure}

\citet{mono-13C-acetone} studied the same 11 rotational transitions with $J \le 4$ 
in the region of 10$-$25~GHz as they did for CH$_3^{13}$C(O)CH$_3$. The lower symmetry 
resulted in 55 transition frequencies, compared to 44 transitions for the symmetrically substituted isotopolog. The resulting spectroscopic parameters were again sufficiently accurate to 
search for transitions with low values of $J$. However, in the case of the asymmetrically 
substituted isotopolog, we did not have an isotopically enriched sample at our disposal. 
Nevertheless, we were able to identify a small number of unblended transitions of 
$^{13}$CH$_3$C(O)CH$_3$ in the acetone spectra recorded in natural isotopic composition. 
The assignments were beyond doubt because of our earlier investigation of an 
isotopically enriched sample. 
Transitions with $K_a=0 \leftarrow 1$ and $K_a=1 \leftarrow 0$ 
are particularly promising because they already overlap at fairly low $J$
(oblate pairing). A typical fingerprint due to internal rotation for these partly 
collapsed transitions consists of three lines. In this case,
transitions of the torsional substate with quantum numbers of
($\sigma_1=0$, $\sigma_2=0$) are observed as single lines (unblended), while lines 
of ($\pm1$, $0$) and (0, $\pm1$) are blended, as are ($\pm1$, $\pm1$) and 
($\pm1$, $\mp1$). This leads to an intensity ratio of $2:(2+2):(1+1) = 1:2:1$, see 
Table~\ref{Table_Int-acetone}. This intensity ratio played a decisive role for assignments,
as can be seen in Fig.~\ref{spectra_13C}.

As a first step, we recorded transitions in the frequency region 36$-$70~GHz and
 assigned transitions with $J'\leftarrow J''=6\leftarrow5$ and $4\leftarrow 3$. 
After these assignments, we assigned transitions in higher frequency 
regions 81$-$130~GHz (setup 1b of Table~\ref{exp_setups} was used) and 244$-$351~GHz
(setup 2a) for $J'$ ranging 
from 8 to 13 and 25 to 36, respectively.
Relative intensities of candidate lines were compared to known unperturbed lines
of the main isotopolog to ensure the assignments because the large number of lines
in the spectral recordings frequently lead to blending.

The sextic centrifugal distortion parameters turned out to be insensitive 
to our dataset and were omitted from the final fit. In addition, 
we fixed $[\Delta_J]_{01}$ to its counterpart $[\Delta_J]_{10}$ and $\epsilon_{02}$ to $\epsilon_{20}$.
The final fit had a weighted standard deviation of 0.93. The resulting 
spectroscopic parameters are given in Table~\ref{spectroscopic-parameters-1-13C}.

\begin{table}
\begin{center}

  \caption{Ground state spectroscopic parameters\tablefootmark{a} of $^{13}$CH$_3$C(O)CH$_3$.}
  \label{spectroscopic-parameters-1-13C}
{\footnotesize
  \begin{tabular}{lr@{}lr@{}l}
 \hline\hline
    Parameter & \multicolumn{2}{c}{Lovas \& Groner} & \multicolumn{2}{c}{This work}  \\
              & \multicolumn{2}{c}{(2006)} &                     \\
     \hline
     \\                          
                           $\rho_1$    &       0&.060591(37)     &       0&.060458~(57)   \\
                           $\rho_2$    &       0&.062047(34)     &       0&.061811~(58)   \\
             $\beta_1$ ~/$^{\circ}$    &      29&.5461(36)       &      29&.5967~(77)     \\
 $180^{\circ}-\beta_2$ ~/$^{\circ}$    &      21&.3035(49)       &      21&.2614~(76)     \\
                          $A$ ~/MHz    &   10083&.0347(11)       &   10083&.03218~(58)    \\
                          $B$ ~/MHz    &    8277&.5070(13)       &    8277&.50617~(44)    \\
                          $C$ ~/MHz    &    4811&.4692(10)       &    4811&.468965~(174)  \\
                 $\Delta_{J}$ ~/kHz    &       4&.62(14)         &       4&.5557~(147)    \\
                $\Delta_{JK}$ ~/kHz    &    $-$2&.60(14)         &    $-$2&.865~(74)      \\
                 $\Delta_{K}$ ~/kHz    &       9&.34(11)         &       9&.456~(60)      \\
                 $\delta_{J}$ ~/kHz    &       1&.901(21)        &       1&.8855~(77)     \\
                 $\delta_{K}$ ~/kHz    &    $-$0&.253(77)        &    $-$0&.226~(31)      \\

                    $\Phi_{J}$ ~/Hz    &       0&.0506046\tablefootmark{b}                              \\
                   $\Phi_{JK}$ ~/Hz    &    $-$0&.336741\tablefootmark{b}                               \\
                   $\Phi_{KJ}$ ~/Hz    &       0&.\tablefootmark{b}                                     \\
                    $\Phi_{K}$ ~/Hz    &       0&.423395\tablefootmark{b}                               \\
                    $\phi_{J}$ ~/Hz    &       0&.0253760\tablefootmark{b}                              \\
                   $\phi_{JK}$ ~/Hz    &    $-$0&.0273291\tablefootmark{b}                              \\
                    $\phi_{K}$ ~/Hz    &    $-$0&.221468\tablefootmark{b}                               \\                                         
                                       
              $\epsilon_{10}$ ~/MHz    &  $-$756&.85(49)         &  $-$759&.25~(81)       \\
              $\epsilon_{01}$ ~/MHz    &  $-$763&.15(41)         &  $-$766&.18~(79)       \\
              $\epsilon_{20}$ ~/MHz    &       0&.766643\tablefootmark{b}      &       0&.574~(45)\tablefootmark{c}  \\              
              $\epsilon_{02}$ ~/MHz    &       0&.766643\tablefootmark{b}      &       0&.574~(45)      \\
              
              $\epsilon_{11}$ ~/MHz    &       1&.049511\tablefootmark{b}     &       1&.0000~(212)    \\
             $\epsilon_{1-1}$ ~/MHz    &       0&.0799732\tablefootmark{b}     &       0&.08630~(143)   \\              
            
           $[A-(B+C)/2]_{01}$ ~/kHz    &      57&.8(26)          &      67&.1~(40)        \\
             $[(B+C)/2]_{01}$ ~/kHz    &   $-$18&.25(37)         &   $-$15&.64~(39)       \\
             $[(B-C)/4]_{01}$ ~/kHz    &    $-$2&.21(18)         &    $-$0&.877(193)      \\
             $[\Delta_J]_{01}$ ~/Hz    &        &                &      10&.053~(91)      \\

           $[A-(B+C)/2]_{10}$ ~/kHz    &      57&.8(26)\tablefootmark{c}      &      66&.2~(34)        \\
             $[(B+C)/2]_{10}$ ~/kHz    &   $-$18&.25(37)\tablefootmark{c}     &   $-$20&.69~(55)       \\
             $[(B-C)/4]_{10}$ ~/kHz    &    $-$2&.21(18)\tablefootmark{c}     &    $-$3&.343~(271)     \\
             $[\Delta_J]_{10}$ ~/Hz    &        &                &      10&.053~(91)\tablefootmark{c}  \\
             
              $[(B-C)/4]_{11}$ ~/Hz    &        &                &   $-$93&.~(33)         \\
                                                                                          \\
                                                                                             
    \hline

no. of transitions fit                 &      55&                &     110&               \\
no. of lines fit                       &      55&                &      72&               \\
 standard deviation$^d$                 &       1&.54             &       0&.93            \\
    \hline

  \end{tabular}\\[2pt]
}
\end{center}
   \tablefoot{
\tablefoottext{a}{Numbers in parentheses are one standard deviation in units of the least significant figures.} 
\tablefoottext{b}{Fixed in the fit.} 
\tablefoottext{c}{Parameter of $(q,q')=(q,0)$ is fixed to its counterpart equivalent parameter with $(0,q).$} 
\tablefoottext{d}{Weighted unitless value for the entire fit.} 
}
\end{table}


\subsubsection{The main isotopic species in its ground vibrational state}
\label{main-iso_GS}

\citet{acetone-12C_groner_2002} studied the main isotopolog quite extensively. 
However, striking discrepancies between the resulting predictions and our ALMA 
spectra of Sgr~B2(N2), detailed in Sect.~\ref{ss:12c}, motivated us to revisit 
the rotational spectroscopy of the main isotopolog. These discrepancies may be 
caused by the insufficient quantum number coverage of the experimental lines 
at higher $J$. The coverage is quite good for transitions without asymmetry 
splitting, that is transitions with high values of $K_a$ (prolate paired) or $K_c$ 
(oblate paired), respectively, even though some of the prolate paired transitions 
did not fit well \citep{acetone-12C_groner_2002}. Transitions with asymmetry 
splitting are largely missing for higher quantum numbers. They are weaker by 
a factor of two and frequently display a more complex internal rotation pattern. 
Another reason may well be that the weighted standard deviation of their fit, 
1.58, is only close to the experimental uncertainty.

We tested parameters with $\omega = -1$ to improve the quality of the fit, 
analogously to our CH$_3^{13}$C(O)CH$_3$ fit. With just two added parameters, 
six other parameters could be omitted and the weighted standard deviation 
of the same dataset still improved to 1.29.
In the fits of \citet{acetone-12C_groner_2002} and our modified fit
of \citet{acetone-12C_groner_2002} (see second and third column of
Table~\ref{spectroscopic-parameters-12C}, respectively),
all transitions were treated as single lines even if they had the same
frequency
since the 2002 version of ERHAM used by \citet{acetone-12C_groner_2002}
did not properly account for blended lines. 

The ERHAM version available nowadays, as our extended version,
is capable of properly treating 
blended lines.\ This frequently leads to lower weighted standard deviations.
We achieved a more decisive
improvement of our model by assigning more than 
1000 transitions of the main isotopolog in the frequency regions of 36$-$70~GHz 
and 82$-$125~GHz. This extension proved to be sufficient for our astronomical 
observations which cover a considerable part of the second frequency region. 
A comparison of the parameters derived by \citet{acetone-12C_groner_2002}, 
the refitted version of their data, and the final parameter values derived 
in this work can be found in Table~\ref{spectroscopic-parameters-12C}. 
A comparison of low-order spectroscopic parameters of the three isotopic 
species of acetone is given in Table~\ref{spectroscopic-parameters-comp}.


\begin{table*}
\begin{center}

  \caption{Ground state spectroscopic parameters\tablefootmark{a} of CH$_3$C(O)CH$_3$.}
  \label{spectroscopic-parameters-12C}
{\footnotesize
  \begin{tabular}{lr@{}lr@{}lr@{}l}
 \hline\hline

    Parameter & \multicolumn{2}{c}{\citet{acetone-12C_groner_2002}} & \multicolumn{2}{c}{modified fit of} & \multicolumn{2}{c}{This work}  \\
     &    &  & \multicolumn{2}{c}{\citet{acetone-12C_groner_2002}} &   \\
     \hline
     \\                          

                       $\rho$          &       0&.0621760~(60)     &      0&.062257~(38)   &       0&.0619535~(120)   \\
               $\beta$ ~/$^{\circ}$    &      25&.8322~(93)        &     26&.196~(48)      &      25&.5065~(76)       \\
                          $A$ ~/MHz    &   10165&.21654~(80)       &  10165&.21755~(64)    &   10165&.217780~(280)    \\
                          $B$ ~/MHz    &    8515&.16477~(65)       &   8515&.16382~(53)    &    8515&.163068~(248)    \\
                          $C$ ~/MHz    &    4910&.19903~(44)       &   4910&.19880~(36)    &    4910&.198777~(209)    \\
                 $\Delta_{J}$ ~/kHz    &       4&.9055~(25)        &      4&.87712~(192)   &       4&.854449~(219)    \\
                $\Delta_{JK}$ ~/kHz    &    $-$3&.620~(17)         &   $-$3&.3984~(103)    &    $-$3&.17067~(60)      \\
                 $\Delta_{K}$ ~/kHz    &      10&.245~(17)         &     10&.0333~(109)    &       9&.79105~(88)      \\
                 $\delta_{J}$ ~/kHz    &       2&.0645~(12)        &      2&.05031~(93)    &       2&.038605~(99)     \\
                 $\delta_{K}$ ~/kHz    &    $-$0&.7393~(56)        &   $-$0&.68451~(285)   &    $-$0&.60720~(39)      \\
                   $\Phi_{J}$ ~/mHz    &      50&.6~(34)           &     30&.59~(166)      &       6&.505~(71)        \\
                  $\Phi_{JK}$ ~/mHz    &  $-$337&.~(20)            & $-$202&.8~(66)        &   $-$32&.02~(31)         \\
                  $\Phi_{KJ}$ ~/mHz    &       0&.                 &      0&.              &      43&.45~(114)        \\
                   $\Phi_{K}$ ~/mHz    &     423&.~(20)            &    279&.1~(51)        &      20&.91~(136)        \\
                   $\phi_{J}$ ~/mHz    &      25&.4~(17)           &     15&.36~(83)       &       3&.225~(34)        \\
                  $\phi_{JK}$ ~/mHz    &   $-$27&.3~(41)           &      0&.              &      39&.595~(166)       \\
                   $\phi_{K}$ ~/mHz    &  $-$221&.5~(83)           & $-$168&.70~(295)      &   $-$76&.16~(58)         \\
                   
              $\epsilon_{10}$ ~/MHz    &  $-$763&.198~(62)         & $-$764&.574~(178)     &  $-$764&.737~(38)        \\
              $\epsilon_{20}$ ~/MHz    &       0&.767~(13)         &      0&.7851~(100)    &       0&.77490~(283)     \\
              $\epsilon_{11}$ ~/MHz    &       1&.050~(43)         &      1&.092~(33)      &       1&.0902~(95)       \\
             $\epsilon_{1-1}$ ~/MHz    &       0&.0800~(83)        &      0&.0800~(61)     &       0&.07346~(262)     \\
           
           $[A-(B+C)/2]_{10}$ ~/kHz    &      55&.07~(64)          &     83&.86~(192)      &      49&.83~(46)         \\
             $[(B+C)/2]_{10}$ ~/kHz    &   $-$21&.16~(56)          &  $-$39&.66~(274)      &    $-$3&.425~(262)       \\
             $[(B-C)/4]_{10}$ ~/kHz    &    $-$3&.40~(27)          &  $-$12&.54~(138)      &       5&.536~(129)       \\

             $[\Delta_J]_{10}$ ~/Hz    &      39&.06~(34)          &     30&.01~(159)      &      41&.66~(63)         \\             
          $[\Delta_{JK}]_{10}$ ~/Hz    &   $-$99&.8~(17)           &  $-$84&.92~(295)      &  $-$121&.52~(172)        \\             
           $[\Delta_{K}]_{10}$ ~/Hz    &      73&.7~(17)           &     54&.64~(259)      &      78&.00~(128)        \\
           $[\delta_{J}]_{10}$ ~/Hz    &      19&.60~(18)          &     15&.15~(80)       &       2&.429~(68)        \\           
           $[\delta_{K}]_{10}$ ~/Hz    &   $-$34&.27~(98)          &  $-$32&.37~(87)       &   $-$19&.10~(43)         \\
                  $[d_2]_{10}$ ~/Hz    &        &                  &       &               &   $-$18&.54~(33)         \\

              $[\Phi_J]_{10}$ ~/mHz    &        &                  &       &               &    $-$9&.268~(286)       \\
           $[\Phi_{JK}]_{10}$ ~/mHz    &        &                  &       &               &      25&.71~(67)         \\
           $[\Phi_{KJ}]_{10}$ ~/mHz    &        &                  &       &               &   $-$24&.59~(60)         \\
            $[\Phi_{K}]_{10}$ ~/mHz    &        &                  &       &               &       9&.13~(70)         \\

            $[\phi_{J}]_{10}$ ~/mHz    &        &                  &       &               &    $-$0&.684~(39))       \\           
           $[\phi_{JK}]_{10}$ ~/mHz    &        &                  &       &               &       4&.034~(105)       \\
            $[\phi_{K}]_{10}$ ~/mHz    &        &                  &       &               &    $-$1&.90~(37)         \\
                 $[h_2]_{10}$ ~/mHz    &        &                  &       &               &       4&.255~(145)       \\                                                                                                
                 $[h_3]_{10}$ ~/mHz    &        &                  &       &               &       0&.311~(39)        \\
   
           $[B^-_{010}]_{10}$ ~/kHz    &        &                  &       &               &      103&.3~(213)        \\       
            $[B^-_{210}]_{10}$ ~/Hz    &        &                  & $-$292&.~(37)         &   $-$273&.1~(92)         \\     
           $[B^-_{001}]_{10}$ ~/kHz    &        &                  &    340&.~(44)         &   $-$292&.90~(280)       \\
           $[B^-_{012}]_{10}$ ~/kHz    &        &                  &       &               &     $-$1&.3420~(271)     \\
           $[B^-_{212}]_{10}$ ~/mHz    &        &                  &       &               &      343&.9~(78)         \\

          $[A-(B+C)/2]_{1-1}$ ~/kHz    &       1&.62~(25)          &       &               &         &                \\
            $[(B+C)/2]_{1-1}$ ~/kHz    &    $-$1&.43~(18)          &       &               &         &                \\
            $[(B-C)/4]_{1-1}$ ~/kHz    &    $-$0&.475~(73)         &       &               &         &                \\             
             
           $[A-(B+C)/2]_{20}$ ~/kHz    &       0&.87~(21)          &       &               &         &                \\
             $[(B+C)/2]_{20}$ ~/kHz    &    $-$0&.31~(13)          &       &               &         &                \\    
                                                                                                                      \\
    \hline

 no. of transitions fit                &    1002&                  &   1002&               &     2181&                \\
 no. of lines fit                      &    1002&\tablefootmark{b}              &   1002&\tablefootmark{b}           &     1862&                \\
 standard deviation\tablefootmark{c}                         &       1&.58               &      1&.29            &        0&.95             \\
    \hline   
  \end{tabular}\\[2pt]
}
\end{center}
   \tablefoot{
\tablefoottext{a}{Numbers in parentheses are one standard deviation in 
units of the least significant figures.} 
\tablefoottext{b}{Blended transitions are fit as single lines.} 
\tablefoottext{c}{Weighted unitless standard  value for the entire fit.} 
}
\end{table*}


\begin{table*}
\begin{center}

  \caption{Selected ground state spectroscopic parameters\tablefootmark{a} of this work
  of CH$_3$C(O)CH$_3$, CH$_3^{13}$C(O)CH$_3$, and $^{13}$CH$_3$C(O)CH$_3$.}
  \label{spectroscopic-parameters-comp}
{\footnotesize
  \begin{tabular}{lr@{}lr@{}lr@{}l}
 \hline\hline

 Parameter &  \multicolumn{2}{c}{CH$_3$C(O)CH$_3$} &  \multicolumn{2}{c}{CH$_3^{13}$C(O)CH$_3$} &  \multicolumn{2}{c}{$^{13}$CH$_3$C(O)CH$_3$} \\
                   
     \hline
     \\          

                    $\rho_1$              &       0&.0619535~(120)        &       0&.0614858~(41)   &      0&.060458~(57)   \\
                    $\rho_2$              &        &                      &        &                &      0&.061811~(58)   \\                    
 $\beta_1$ ~/$^{\circ}$                   &       25&.5065~(76)           &      25&.7140~(31)      &     29&.5967~(77)     \\
 $180^{\circ}-\beta_2$ ~/$^{\circ}$       &         &                     &        &                &     21&.2614~(76)     \\ 
                             $A$ ~/MHz    &       10165&.217780~(280)     &   10164&.005782~(145)   &  10083&.03218~(58)    \\
                             $B$ ~/MHz    &        8515&.163068~(248)     &    8516&.083092~(104)   &   8277&.50617~(44)    \\
                             $C$ ~/MHz    &        4910&.198777~(209)     &    4910&.235399~(107)   &   4811&.468965~(174)  \\
                    $\Delta_{J}$ ~/kHz    &           4&.854449~(219)     &       4&.852046~(73)    &      4&.5557~(147)    \\
                   $\Delta_{JK}$ ~/kHz    &        $-$3&.17067~(60)       &    $-$3&.192449~(194)   &   $-$2&.865~(74)      \\
                    $\Delta_{K}$ ~/kHz    &           9&.79105~(88)       &       9&.85055~(33)     &      9&.456~(60)      \\
                    $\delta_{J}$ ~/kHz    &           2&.038605~(99)      &       2&.0381002~(265)  &      1&.8855~(77)     \\
                    $\delta_{K}$ ~/kHz    &        $-$0&.60720~(39)       &    $-$0&.619619~(127)   &   $-$0&.226~(31)      \\

                 $\epsilon_{10}$ ~/MHz    &       $-$764&.737~(38)        &  $-$763&.9260~(53)      & $-$759&.25~(81)       \\
                 $\epsilon_{01}$ ~/MHz    &             &                 &        &                & $-$766&.18~(79)       \\                 
                 $\epsilon_{20}$ ~/MHz    &       0&.77490~(283)          &       0&.76612~(173)    &      0&.574~(45)\tablefootmark{b}  \\
                 $\epsilon_{02}$ ~/MHz    &        &                      &        &                &      0&.574~(45)      \\                 
                 $\epsilon_{11}$ ~/MHz    &       1&.0902~(95)            &       1&.1059~(61)      &      1&.0000~(212)    \\
                $\epsilon_{1-1}$ ~/MHz    &       0&.07346~(262)          &       0&.08853~(176)    &      0&.08630~(143)   \\

    \hline

 no. of transitions fit                   &    2181&                      &    9715&                &     110&       \\
 no. of lines fit                         &    1862&                      &    5870&                &      72&     \\
 standard deviation\tablefootmark{c}      &       0&.95                   &       1&.29             &   0&.93     \\

    \hline    
  \end{tabular}\\[2pt]
}
\end{center}
   \tablefoot{
\tablefoottext{a}{Numbers in parentheses are one standard deviation in units of the least significant figures.} 
\tablefoottext{b}{Parameter of $(q,q')=(q,0)$ is fixed to its counterpart equivalent parameter with $(0,q).$} 
\tablefoottext{c}{Weighted unitless value for the entire fit.} 
}
\end{table*}


\subsubsection{The main isotopic species in its first two torsionally excited states}
\label{main-iso_ES} 

Clear indications in the astronomical emission spectra of two torsionally excited states 
of the main species, $\varv _{12} = 1$ and $\varv _{17} = 1$, made it indispensable 
to extend the line lists to ensure clear identification of lines within them. 
Fitting and modeling the first two torsionally excited states was considerably more 
challenging. For both states, the overall aim in modeling the spectra was to derive 
models that fit our data close to experimental uncertainty to simply allow for 
astronomical detection of these species. Nevertheless, the initial line lists 
for $\varv _{12} = 1$ and $\varv _{17} = 1$, consisting of around 400 and 600 lines from 
\citet{acetone_v12eq1_2006} and \citet{acetone_v17eq1_2008}, respectively, were 
extended by almost 1300 and more than 600 lines, respectively.

Attempts to improve the fits of the two excited torsional states of the main isotopic 
species were only partially successful, the weighted standard deviation values were 1.36 and 1.62 for 
$\varv _{12} = 1$ and $\varv _{17} = 1$, respectively. Therefore, only selected parameters 
of these states are listed and compared to values of the ground vibrational state of 
acetone-{$^{12}$C} in Table~\ref{spectroscopic-parameters-12C-VES}. 
The full list of parameters can be found in Tables~\ref{A:spectroscopic-parameters-12C-v12} 
and \ref{A:spectroscopic-parameters-12C-v24}, for $\varv _{12} = 1$ and 
$\varv _{17} = 1$, respectively.


\begin{table*}
\begin{center}

  \caption{Selected\tablefootmark{a} spectroscopic parameters\tablefootmark{b} of the 
  torsionally excited states $\varv_{12}=1$ and $\varv_{17}=1$ of CH$_3$C(O)CH$_3$ 
  in comparison to the ground state values ($\varv=0$).}
  \label{spectroscopic-parameters-12C-VES}
  \begin{tabular}{lr@{}lr@{}lr@{}l}
  \hline

    Parameter & \multicolumn{2}{c}{$\varv=0$} & \multicolumn{2}{c}{$\varv_{12}=1$} & \multicolumn{2}{c}{$\varv_{17}=1$}  \\
     \hline
     \\                          

                  $\rho$           &         0&.0619535(120)     &        0&.0624819(165)     &         0&.062788(80)   \\
       $\beta$ ~/$^{\circ}$    &        25&.5065(76)         &       25&.6443(66)         &        27&.100(45)      \\
                  $A$ ~/MHz    &     10165&.217780(280)      &    10177&.0066(33)         &     10191&.123(98)      \\
                  $B$ ~/MHz    &      8515&.163068(248)      &     8502&.97353(212)       &      8480&.859(81)      \\
                  $C$ ~/MHz    &      4910&.198777(209)      &     4910&.34514(154)       &      4910&.06430(229)   \\
         $\Delta_{J}$ ~/kHz    &         4&.854449(219)      &        4&.9566(102)        &         7&.812(109)     \\
        $\Delta_{JK}$ ~/kHz    &      $-$3&.17067(60)        &        0&.460(41)          &         2&.84(118)      \\
         $\Delta_{K}$ ~/kHz    &         9&.79105(88)        &        1&.631(91)          &     $-$37&.67(191)      \\
         $\delta_{J}$ ~/kHz    &         2&.038605(99)       &        2&.0942(54)         &         3&.535(54)      \\
         $\delta_{K}$ ~/kHz    &      $-$0&.60720(39)        &        0&.6937(212)        &         2&.57(46)       \\
                   
      $\epsilon_{10}$ ~/MHz    &    $-$764&.737(38)          &     5614&.30(201)          &     12970&.0(287)       \\
      $\epsilon_{20}$ ~/MHz    &         0&.77490(283)       &       19&.329(108)         &       458&.9(32)        \\
      $\epsilon_{30}$ ~/MHz    &          &                  &         &                  &        21&.24(32)       \\
      $\epsilon_{11}$ ~/MHz    &         1&.0902(95)         &       67&.223(61)          &       102&.91(295)      \\
     $\epsilon_{1-1}$ ~/MHz    &         0&.07346(262)       &      100&.977(60)          &     $-$30&.21(74)       \\
                                                                                                                \\

    \hline

no. of transitions fit      &  2181&        & 1621&      & 939   \\
no. of lines fit            &  1862&        & 1298&      & 719   \\
 standard deviation\tablefootmark{c}      &     0&.95     &    1&.36   & 1&.62 \\
    \hline   
  \end{tabular}\\[2pt]
\end{center}
   \tablefoot{
\tablefoottext{a}{The full list of spectroscopic parameters is in the Appendix in 
 Tables~\ref{A:spectroscopic-parameters-12C-v12} and \ref{A:spectroscopic-parameters-12C-v24}}. 
\tablefoottext{b}{Numbers in parentheses are one standard deviation in units of the least significant figures.} 
\tablefoottext{c}{Weighted unitless value for the entire fit.} 
}
\end{table*}


\subsection{Radioastronomical observations}
\label{obs-astro}

We used the EMoCA spectral line survey performed toward the high-mass star 
forming region Sagittarius~B2(N) with the Atacama Large 
Millimeter/submillimeter Array (ALMA) to search for the 
$^{13}$C isotopologs of acetone and for transitions of the $^{12}$C isotopolog 
in its torsionally excited states $\varv_{12}=1$ and $\varv_{17}=1$. Details 
about the observations, data reduction, and the method used to identify 
the detected lines and derive column densities can be found in 
\citet{2x13C-MeCN_det_2016}. In short, the survey covers the frequency range from 
84.1~GHz to 114.4~GHz with a spectral resolution of 488.3~kHz (1.7 to 
1.3~km~s$^{-1}$). It was performed with five frequency tunings called S1 to S5. 
The median angular resolution is 1.6$\arcsec$. In this work we focus on the peak 
position of the hot molecular core Sgr~B2(N2) with J2000 equatorial coordinates 
($17^{\rm h}47^{\rm m}19.86^{\rm s}$, $-28^\circ22\arcmin13.4\arcsec$).

\subsubsection{CH$_3$C(O)CH$_3$}
\label{ss:12c}

Before searching for transitions of the $^{13}$C isotopologs of acetone, we 
first modeled the emission of the main isotopolog in the EMoCA survey of Sgr~B2(N2). 
As described in \citet{2x13C-MeCN_det_2016}, we proceeded in an iterative way 
using Weeds \citep[][]{Maret11} under the local thermodynamic equilibrium (LTE) 
approximation. Each species was modeled with the following five free parameters: size of 
the emission region, rotational temperature, column density, line width, 
and velocity offset with respect to the systemic velocity of the source. 
We modeled the emission of acetone on top of the contribution of all the molecules 
we already identified so far in this survey. 
This allowed us to spot 26 spectral lines of acetone that are not contaminated 
by the emission of other species. We used these transitions to fit the size of 
the acetone emission in the interferometric maps toward Sgr~B2(N2). We derived 
a median size (FWHM) of 1.2$\arcsec$. The line width and velocity offset were 
derived directly from these uncontaminated transitions.

\begin{figure}
\centerline{\resizebox{0.95\hsize}{!}{\includegraphics[angle=0]{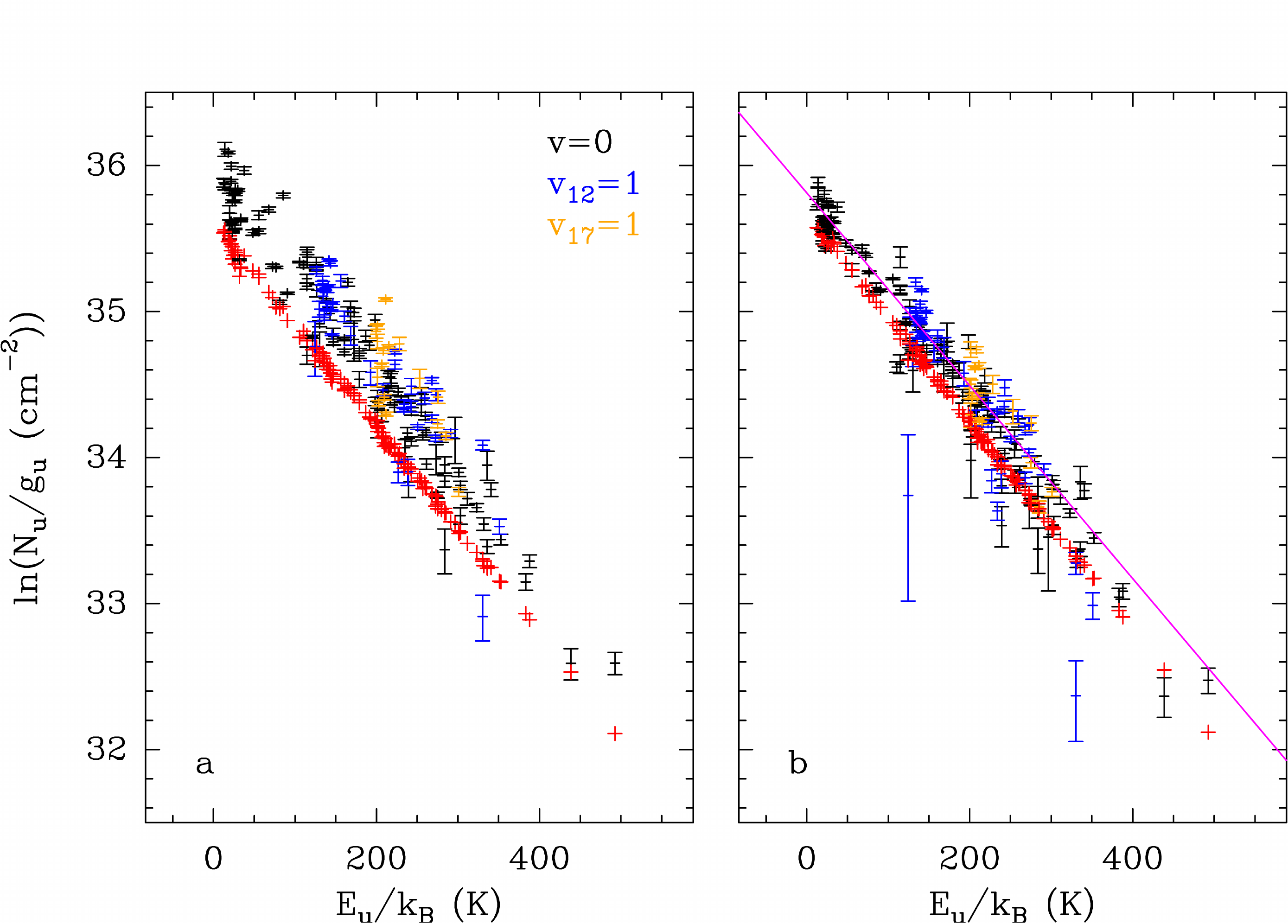}}}
\caption{Population diagram of CH$_3$C(O)CH$_3$ toward Sgr~B2(N2). The 
observed datapoints are shown in black, blue, and orange for the
$\varv=0$, $\varv_{12}=1$, and $\varv_{17}=1$ states, respectively,
while the synthetic populations are 
shown in red. No correction is applied in panel \textbf{a}. 
In panel \textbf{b}, the optical depth correction was applied to both the 
observed and synthetic populations and the contamination by all other 
species included in the full model was removed from the observed 
datapoints. The purple line is a linear fit to the observed populations (in 
linear-logarithmic space).
}
\label{f:popdiag_ch3coch3}
\end{figure}

With these parameters and a first guess of the rotational temperature, we 
computed synthetic spectra for $\varv=0$, $\varv_{12}=1$, and $\varv_{17}=1$, 
which allowed us to identify rotational transitions of both torsionally excited 
states in the EMoCA spectrum of Sgr~B2(N2). We used all the transitions 
of these states and the vibrational ground state that were not 
contaminated too much by other species to build a population diagram. This diagram is 
shown in Fig.~\ref{f:popdiag_ch3coch3}. It contains 110, 40, and 17 spectral
lines of $\varv=0$, $\varv_{12}=1$, and $\varv_{17}=1$, respectively. 
Figure~\ref{f:popdiag_ch3coch3}b shows the population diagram after correcting 
the integrated intensities for the line optical depth and contamination 
due to the other identified species that are included in our full model. A fit 
to this diagram yields a rotational temperature of $152 \pm 4$~K.
This temperature is a fair indication of the true 
rotational temperature of the molecule. However, its uncertainty is likely 
underestimated because the dispersion of the datapoints in 
Fig.~\ref{f:popdiag_ch3coch3}b is dominated by residual contamination from 
unidentified species while the errors reported on the individual datapoints and
used for the weighted fit are purely statistical. For comparison purposes, a fit 
limited to the transitions in the vibrational ground state only yields a 
temperature of $147 \pm 5$~K. In the following, we used a temperature of 140~K, 
which is still close to the fitted value, to model the spectrum of acetone. 

\begin{table*}[!ht]
 \begin{center}
 \caption{
 Parameters of our best-fit LTE model of acetone and its $^{13}$C isotopologs toward Sgr~B2(N2).
}
 \label{t:coldens}
 \vspace*{-1.2ex}
 \begin{tabular}{lcrccccccr}
 \hline\hline
 \multicolumn{1}{c}{Molecule} & \multicolumn{1}{c}{Status\tablefootmark{a}} & \multicolumn{1}{c}{$N_{\rm det}$\tablefootmark{b}} & \multicolumn{1}{c}{Size\tablefootmark{c}} & \multicolumn{1}{c}{$T_{\mathrm{rot}}$\tablefootmark{d}} & \multicolumn{1}{c}{$N$\tablefootmark{e}} & \multicolumn{1}{c}{$F_{\rm vib}$\tablefootmark{f}} & \multicolumn{1}{c}{$\Delta V$\tablefootmark{g}} & \multicolumn{1}{c}{$V_{\mathrm{off}}$\tablefootmark{h}} & \multicolumn{1}{c}{$\frac{N_{\rm ref}}{N}$\tablefootmark{i}} \\ 
  & & & \multicolumn{1}{c}{($''$)} & \multicolumn{1}{c}{(K)} & \multicolumn{1}{c}{(cm$^{-2}$)} & & \multicolumn{1}{c}{(km~s$^{-1}$)} & \multicolumn{1}{c}{(km~s$^{-1}$)} & \\ 
 \hline
 CH$_3$C(O)CH$_3$, $\varv=0$$^\star$ & d & 122 &  1.2 &  140 &  4.0 (17) & 1.00 & 5.0 & 0.0 &       1 \\ 
 \hspace*{13ex} $\varv_{12}=1$ & d & 33 &  1.2 &  140 &  4.0 (17) & 1.00 & 5.0 & 0.0 &       1 \\ 
 \hspace*{13ex} $\varv_{17}=1$ & d & 13 &  1.2 &  140 &  4.0 (17) & 1.00 & 5.0 & 0.0 &       1 \\ 
 $^{13}$CH$_3$C(O)CH$_3$, $\varv=0$ & t & 1 &  1.2 &  140 &  3.0 (16) & 1.00 & 5.0 & 0.0 &      13 \\ 
 CH$_3$$^{13}$C(O)CH$_3$, $\varv=0$ & n & 0 &  1.2 &  140 & $<$  1.5 (16) & 1.00 & 5.0 & 0.0 & $>$      27 \\ 
\hline 
 \end{tabular}
 \end{center}
 \vspace*{-2.5ex}
 \tablefoot{
 \tablefoottext{a}{d: detection, t: tentative detection, n: nondetection.}
 \tablefoottext{b}{Number of detected lines \citep[conservative estimate, see Sect.~3 of][]{2x13C-MeCN_det_2016}. One line of a given species may mean a group of transitions of that species that are blended together.}
 \tablefoottext{c}{Source diameter (\textit{FWHM}).}
 \tablefoottext{d}{Rotational temperature.}
 \tablefoottext{e}{Total column density of the molecule. $X$ ($Y$) means $X \times 10^Y$. An identical value for all listed vibrational states of a molecule means that LTE is an adequate description of the vibrational excitation.}
 \tablefoottext{f}{Correction factor that was applied to the column density to account for the contribution of vibrationally excited states, in the cases where this contribution was not included in the partition function of the spectroscopic predictions.}
 \tablefoottext{g}{Linewidth (\textit{FWHM}).}
 \tablefoottext{h}{Velocity offset with respect to the assumed systemic velocity of Sgr~B2(N2), $V_{\mathrm{sys}} = 74$ km~s$^{-1}$.}
 \tablefoottext{i}{Column density ratio, with $N_{\rm ref}$ the column density of the previous reference species marked with a $\star$.}
 }
 \end{table*}

The last parameter to adjust is the column density. We obtained a value of 
$4 \times 10^{17}$~cm$^{-2}$ by fitting the observed spectrum (see 
Table~\ref{t:coldens}). The best-fit synthetic spectra of $\varv=0$, 
$\varv_{12}=1$, and $\varv_{17}=1$ are overlaid on the EMoCA spectrum of 
Sgr~B2(N2) in Figs.~\ref{f:spec_ch3coch3_ve0}, \ref{f:spec_ch3coch3_v12e1}, 
and \ref{f:spec_ch3coch3_v24e1}, respectively. We also display the full model
that contains the contributions of all species identified so far. We counted 
122, 33, and 13 transitions or groups of transitions of acetone that are 
clearly detected in its vibrational ground state $\varv=0$ and its 
torsionally excited states $\varv_{12}=1$ and $\varv_{17}=1$, respectively, 
with little contamination from other species.

\subsubsection{$^{13}$CH$_3$C(O)CH$_3$ and CH$_3^{13}$C(O)CH$_3$}
\label{ss:13c}

We used the parameters derived for the main isotopolog of acetone to search 
for transitions of the $^{13}$C-substituted isotopologs $^{13}$CH$_3$C(O)CH$_3$
and CH$_3^{13}$C(O)CH$_3$. On the one hand, we find tentative evidence for 
the presence of $^{13}$CH$_3$C(O)CH$_3$ in Sgr~B2(N2). The transitions covered 
by our survey are shown in Fig.~\ref{f:spec_ch3coch3_13c1_ve0}. Only one 
transition (at 110.131~GHz) is clearly detected with little contamination from 
other species, but several transitions contribute significantly to the 
detected signal as well. The discrepancy at 108.025~GHz is at the 2$\sigma$ 
level, so it is probably not significant. It may be due to a slight 
overestimate of the baseline level. The synthetic spectrum shown in 
Fig.~\ref{f:spec_ch3coch3_13c1_ve0} was computed with a column density of 
$3.0 \times 10^{16}$~cm$^{-2}$ (see Table~\ref{t:coldens}), which is 13 times 
lower than the column density of the main isotopolog. Accounting for the 
fact that $^{13}$CH$_3$C(O)CH$_3$ contains two equivalent methyl groups, this 
yields a $^{12}$C/$^{13}$C isotopic ratio of 27 for acetone, which is in agreement with 
the isotopic ratio derived for methanol and ethanol in this source 
\citep[25, see][]{13C-EtOH_det_2016}. This gives us confidence that the assignment of
the transition at 110.131~GHz to $^{13}$CH$_3$C(O)CH$_3$ is robust.

On the other hand, we did not detect any transition from CH$_3^{13}$C(O)CH$_3$.
In Fig.~\ref{f:spec_ch3coch3_13c2_ve0}, we show a synthetic spectrum computed
with a column density of $1.5 \times 10^{16}$~cm$^{-2}$, that is half of the
column density of $^{13}$CH$_3$C(O)CH$_3$ as expected if the two species are 
drawn from the same parent $^{12}$C/$^{13}$C isotopic ratio.
Fractionation of $^{13}$C  may lead to a different column density
as discussed in \cite{13C-DME_2013} for the case of dimethyl ether. However,
so far, we have no indications of $^{13}$C fractionation of complex organic
molecules in Sgr~B2(N2),
see also \citet{2x13C-MeCN_det_2016} and \citet{13C-EtOH_det_2016}.
The synthetic spectrum is consistent with the observed one, suggesting that
this isotopolog may be present in Sgr~B2(N2) at this column density level.
However, we can not exclude a higher or lower column density level
and given that no single transition is individually detected, we report a
nondetection in Table~\ref{t:coldens}.

\section{Discussion}

\subsection{Laboratory spectroscopy}
\label{lab-discussion}

We were able to extend the experimental line list of CH$_3^{13}$C(O)CH$_3$ 
thanks to an isotopically enriched sample. The high values of $J$ and $K$ 
of up to 92 and 45, respectively, ensure that observations of this symmetric 
isotopolog in its ground vibrational state are not limited by laboratory 
spectroscopy. In comparison, Groner et al. (2002) assigned lines of the 
main isotopolog CH$_3$C(O)CH$_3$ up to $J=60$.

An isotopically enriched sample for 
$^{13}$CH$_3$C(O)CH$_3$ is not yet available. Therefore, our improvements were more modest for 
this isotopolog. Nevertheless, the data were sufficient to identify this 
species in our ALMA data tentatively.

We doubled the number of assigned transitions of the main isotopolog over the 
course of our investigations and obtained a greatly improved spectroscopic 
parameter set that reproduced our astronomical observations well. 
One explanation in comparison to the model of \citet{acetone-12C_groner_2002} 
may be that transitions of complex molecules need extensive quantum number 
coverage in the assignments that is preferably over the full frequency range 
for proper modeling. This was also observed by some of the 
co-authors in the case of propanal \citep{propanal2017}.

As mentioned in Sect.~\ref{spec-props}, substitution of the carbonyl-C by $^{13}$C 
retains the symmetry. In addition, Table~\ref{spectroscopic-parameters-comp} 
demonstrates that the low-order spectroscopic parameters are, in fact, very similar 
to those of the main isotopolog.\ This is expected because of the proximity of
the carbonyl-C to the center of mass of the molecule. The asymmetrically 
substituted $^{13}$C isotopolog displays pronounced differences in the rotational 
and quartic centrifugal distortion parameters. The tunneling parameters differ 
slightly, and the averages for the two rotors are usually quite close to 
the values of the other two isotopologs; somewhat larger deviations are probably 
caused by the experimental line list that is still fairly small.

Progress on the torsionally excited states of CH$_3$C(O)CH$_3$ was not 
entirely satisfactorily despite the greatly extended line lists. 
Nevertheless, the spectroscopic parameters were sufficient for assignment of 
transitions within these excited states in our astronomical data. However, 
striking deviations occur between some observed and calculated transition 
frequencies, suggesting that these models are still not satisfactory. This is 
reflected in the weighted standard deviations of the fits of 1.36 and 1.62, 
respectively, and by the uncommon selection and quite large number of 
spectroscopic parameters, see Tables \ref{A:spectroscopic-parameters-12C-v12} 
and \ref{A:spectroscopic-parameters-12C-v24}. 
\citet{acetone-fitting_2013} explained the failure of ERHAM to fit 
the excited torsional states of acetone satisfactorily by the amount of 
torsion-torsion interaction that is not modeled in single state fits. 
These interactions are treated satisfactorily in the
PAM\_C2v\_2tops program \citep{acetone-fitting_2013}.
\citet{acetone-12C_2016} fit 12128 microwave and 7 FIR line frequencies
in a joined fit of of $\varv=0$, $\varv_{12}=1$, and $\varv_{17}=1$ of
CH$_3$C(O)CH$_3$ with a
reasonable number of 99 parameters
to a root-mean-square deviation of merely 0.78, highlighting the advantages
of including torsion-torsion interactions.

\subsection{Radioastronomical observations}
\label{astro-discussion}

\begin{figure*}
\centerline{\resizebox{0.95\hsize}{!}{\includegraphics[angle=0]{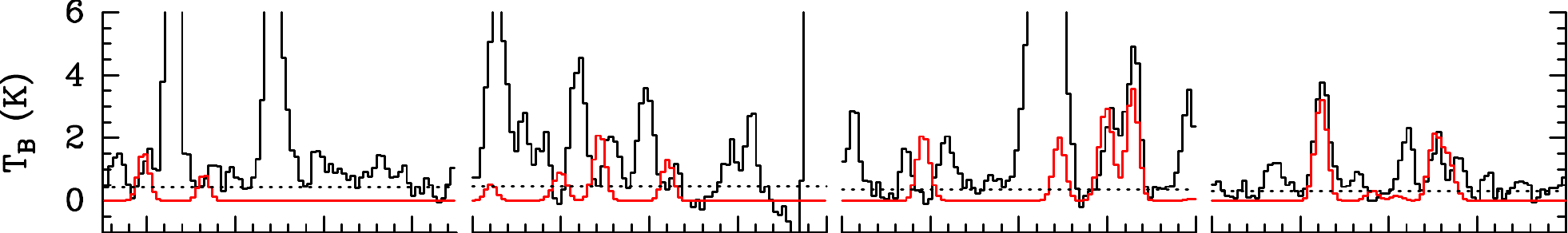}}}
\vspace*{2ex}
\centerline{\resizebox{0.95\hsize}{!}{\includegraphics[angle=0]{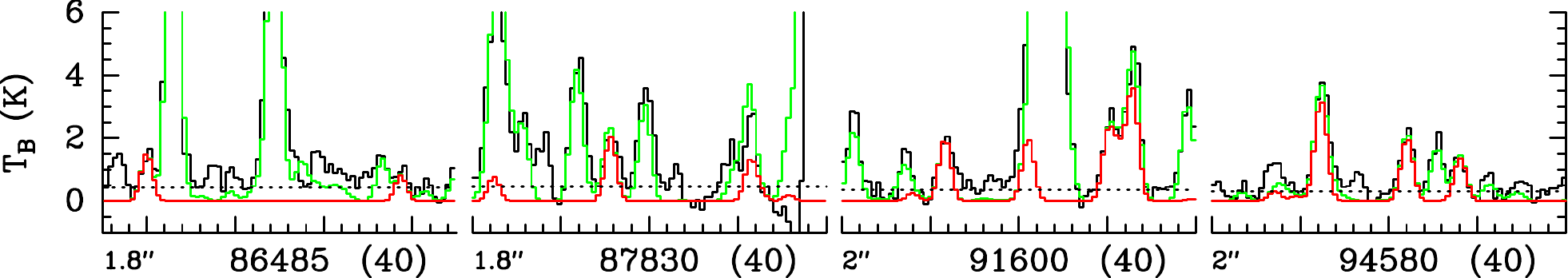}}}
\caption{Examples of lines of acetone covered by EMoCA toward Sgr~B2(N2). The ALMA spectrum is shown in black in all panels. The red 
spectrum shows the synthetic spectrum of acetone computed with the JPL entry (tag 58003, version 1) 
in the top row and with the new predictions presented in this work in the 
bottom row. The synthetic spectrum of all molecules identified so far, 
including acetone, is shown in green. The central frequency and width of the 
frequency range (in parentheses) are indicated at the bottom in MHz. The 
angular resolution (HPBW) is also indicated. The y-axis is labeled in 
brightness temperature units. The dotted line indicates the $3\sigma$ noise 
level. The new predictions match the observed spectrum much better.}
\label{f:comp_old_new}
\end{figure*}

The new spectroscopic predictions produced over the course of this work yield much better agreement with the ALMA spectrum than what could previously be 
achieved with the predictions available in the JPL catalog 
\citep{Pickett98}, which were contributed by B.~J.~Drouin in 2008 by
recalculating the data from \cite{acetone-12C_groner_2002} and references therein.
Figure~\ref{f:comp_old_new} compares the synthetic 
spectra computed with the same LTE parameters for the vibrational ground state
in both cases for four selected frequency ranges. While prominent 
discrepancies were present between the JPL predictions and the observed 
spectrum, for instance at 87.820~GHz, 87.824~GHz, or 91.592~GHz, the synthetic 
spectrum produced with the new predictions agrees perfectly well with the 
observed spectrum. This confirms that the new predictions of the vibrational 
ground state of acetone are more reliable, at least in the 3~mm wavelength range.

The agreement between the synthetic spectra of $\varv_{12}=1$ and $\varv_{17}=1$
with the EMoCA spectrum of Sgr~B2(N2) is good (see Figs.~\ref{f:spec_ch3coch3_v12e1} 
and \ref{f:spec_ch3coch3_v24e1}). Still, we notice discrepancies larger than 
$3\sigma$ and smaller than $\sim5\sigma$ at the following few frequencies: 
93.254~GHz, 98.489~GHz, 103.828~GHz, and 108.857~GHz for 
$\varv_{12}=1$, and 95.464~GHz and 111.541~GHz for $\varv_{17}=1$. Possible 
reasons for these discrepancies are inaccuracies of the predicted frequencies 
of acetone (e.g., at 93.254~GHz, 98.489~GHz, 103.828~GHz, and 95.464~GHz with formal 
uncertainties of the predictions larger than 100~kHz), 
uncertainties as to the true level of the baseline, and inaccuracies of the predicted 
frequencies of contaminating species in the cases where the discrepancy does 
not directly arise from the acetone spectrum but from the full model. All in 
all, these small discrepancies that may be solved with higher-accuracy 
predictions, as mentioned in Sect. \ref{main-iso_ES} and discussed in Sect.
\ref{lab-discussion}, do not affect the reliability of our identifications
of both torsionally excited states.

\section{Conclusion}
\label{conclusion}

The main isotopic species of acetone, CH$_3$C(O)CH$_3$, and CH$_3^{13}$C(O)CH$_3$
are described by robust models, allowing for proper modeling of these species 
in astronomical sources. The predicted transition frequencies by our derived 
models for CH$_3^{13}$C(O)CH$_3$ and CH$_3$C(O)CH$_3$ are reliable 
into the terahertz region. 
Therefore, these models can be of high interest for the analyses of 
hot-core and hot-corino surveys performed with ALMA in all its bands. 
We intend to revisit the spectrum of $^{13}$CH$_3$C(O)CH$_3$ with an enriched sample.
The model derived for $^{13}$CH$_3$C(O)CH$_3$ with a standard deviation of 0.93  
is a good and robust cornerstone for continuing the study of this species in 
the millimeter- and submillimeter-wavelength region in the future.
Predictions of transitions involving quantum numbers where $J=K_c$ may 
be reliable up to 500~GHz or $J\approx 50$. Transition frequencies with 
lower $K_c$ should be viewed with increased caution. Thus far, the vibrational 
ground states of these three isotopologs can be treated as isolated states.

On the other hand, the developer of ERHAM, P. Groner and his co-authors stated in 
a recent paper "that the ERHAM code is unable to fit the $\varv_{17}=1$ state 
of acetone without accounting for the interactions with other vibrational states" 
\citep{acetone-v17_2019} as was explained by \citet{acetone-fitting_2013}. 
The same conclusion may be drawn by looking at the results of this work for the 
vibrationally excited states $\varv_{12}=1$ and $\varv_{17}=1$. Modeling of 
these states requires including interactions between different vibrational states. 
A more robust model for the torsionally excited states is derived by 
\cite{acetone-12C_2016} using the PAM\_C2v\_2tops program 
\citep{acetone-fitting_2013}. This simultaneously fits the ground state and 
both torsionally excited states; however, no linelist is publicly available. 
Such an approach may also be applied beneficially to the torsionally 
excited states of the $^{13}$C-species in the future. 
It is especially important to note that for the symmetric isotopolog CH$_3^{13}$C(O)CH$_3$, 
transition frequencies of excited states with a symmetric motion around 
the symmetry axis of the molecule are only marginally shifted compared 
to the main species.\ This occurs because the replaced carbon atom lies on this symmetry 
axis, which coincides with the observable rotation of the molecule around 
the $b$-axis.

Predictions of the ground state rotational spectra of the three isotopic species 
of acetone in the study will be available in the catalog section of the 
CDMS\footnote{https://cdms.astro.uni-koeln.de/classic/entries/} \citep{CDMS_3}. 
All fit files and auxiliary files will be available in the data section of the 
CDMS\footnote{https://cdms.astro.uni-koeln.de/classic/predictions/daten/Propanon/}.


\begin{acknowledgements}
M.H.O. and O.Z. are very thankful to Peter Groner for various comments and 
recommendations for using his ERHAM program. 
This paper makes use of the following ALMA data: 
ADS/JAO.ALMA\#2011.0.00017.S, ADS/JAO.ALMA\#2012.1.00012.S. 
ALMA is a partnership of ESO (representing its member states),
NSF (USA) and NINS (Japan), 
together with NRC (Canada), NSC and ASIAA (Taiwan), and KASI (Republic of Korea), 
in cooperation with the Republic of Chile. The Joint ALMA Observatory
is operated by ESO, 
AUI/NRAO, and NAOJ. The interferometric data are available in the ALMA archive at
https://almascience.eso.org/aq/. This work is supported by the
Collaborative Research Centre 956, sub-project B3,
funded by the Deutsche Forschungsgemeinschaft (DFG) – project ID 184018867. 
Our research benefited from NASA's Astrophysics Data System (ADS). 
\end{acknowledgements}



\begin{appendix}
\section{Complementary spectroscopic parameters}
\label{Appendix_parameters}

Tables \ref{A:spectroscopic-parameters-12C-v12} and 
\ref{A:spectroscopic-parameters-12C-v24} show the full
list of spectroscopic parameters of $\upsilon_{12}=1$ and
$\upsilon_{17}=1$ of CH$_3$C(O)CH$_3$, respectively. 
A selection of these parameters is listed in
Table~\ref{spectroscopic-parameters-12C-VES}
and compared to the ground vibrational state of
CH$_3$C(O)CH$_3$. The derived models for
the vibrationally excited states should be viewed with caution.


\begin{table*}
\begin{center}

  \caption{Spectroscopic parameters$^a$ of the vibrationally
excited state $\upsilon_{12}=1$ of CH$_3$C(O)CH$_3$.}
  \label{A:spectroscopic-parameters-12C-v12}
  \footnotesize{
  \begin{tabular}{lr@{}llr@{}l}
  \hline

    Parameter & \multicolumn{2}{c}{$\upsilon_{12}=1$} & Parameter & \multicolumn{2}{c}{$\upsilon_{12}=1$}   \\
     \hline
     \\                          
                           $\rho$     &      0&.0624819(165)      &            $[M_{KJ}]_{10}$ ~/\textmu Hz    &          331&.8(233) \\
               $\beta$ ~/$^{\circ}$    &        25&.6443(66)       &                    $[\Phi_J]_{10}$ ~/Hz    &          2&.243(37) \\
                          $A$ ~/MHz    &     10177&.0066(33)       &                  $[L_{JJK}]_{10}$ ~/mHz    &          3&.586(267) \\
                          $B$ ~/MHz    &     8502&.97353(212)      &            $[M_{JK}]_{10}$ ~/\textmu Hz    &          $-$66&.7(42) \\
                          $C$ ~/MHz    &     4910&.34514(154)      &             $[L_{J}]_{10}$ ~/\textmu Hz    &         $-$234&.4(216) \\
                 $\Delta_{J}$ ~/kHz    &         4&.9566(102)      &                  $[B_{021}]_{10}$ ~/kHz    &           9&.27(39) \\
                $\Delta_{JK}$ ~/kHz    &          0&.460(41)       &                  $[(B-C)/4]_{10}$ ~/kHz    &         $-$463&.5(63) \\
                 $\Delta_{K}$ ~/kHz    &          1&.631(91)       &                  $[B_{012}]_{10}$ ~/kHz    &        $-$142&.82(203) \\
                 $\delta_{J}$ ~/kHz    &         2&.0942(54)       &               $[\delta_{K}]_{10}$ ~/kHz    &         $-$2&.328(54) \\
                 $\delta_{K}$ ~/kHz    &         0&.6937(212)      &             $[m_{K}]_{10}$ ~/\textmu Hz    &          43&.45(181) \\
                   $\Phi_{J}$ ~/mHz    &          120&.5(246)      &                 $[\phi_{JK}]_{10}$ ~/Hz    &          1&.580(108) \\
                  $\Phi_{JK}$ ~//Hz    &         14&.530(198)      &                 $[\phi_{J}]_{10}$ ~/mHz    &          521&.4(97) \\
                  $\Phi_{KJ}$ ~//Hz    &         $-$24&.27(55)     &                   $[l_{JK}]_{10}$ ~/mHz    &         $-$2&.926(139) \\
                   $\Phi_{K}$ ~//Hz    &         $-$14&.68(70)     &           $[m_{JJK}]_{10}$ ~/\textmu Hz    &         $-$2&.075(122) \\
                   $\phi_{J}$ ~/mHz    &           72&.8(127)      &                       $[d_2]_{10}$ ~/Hz    &         $-$186&.6(57) \\
                   $\phi_{JK}$ ~/Hz    &          6&.922(102)      &                   $[B_{024}]_{10}$ ~/Hz    &          2&.807(165) \\
                   $\phi_{K}$ ~/mHz    &           $-$868&(146)    &           $[B_{064}]_{10}$ ~/\textmu Hz    &         $-$124&.6(61) \\
             $\epsilon_{1-1}$ ~/MHz    &        100&.977(60)       &                       $[h_2]_{10}$ ~/Hz    &        $-$1&.1388(191) \\
              $\epsilon_{10}$ ~/GHz    &        5&.61430(201)      &                  $[B_{224}]_{10}$ ~/mHz    &         $-$2&.073(149) \\
              $\epsilon_{11}$ ~/MHz    &         67&.223(61)       &           $[B_{244}]_{10}$ ~/\textmu Hz    &          34&.34(209) \\
              $\epsilon_{20}$ ~/MHz    &         19&.329(108)      &               $[l_2]_{10}$ ~/\textmu Hz    &          133&.3(117) \\
                      $L_{K}$ ~/mHz    &          16&.34(114)      &                      $[h_3]_{10}$ ~/mHz    &         $-$506&.7(95) \\
                    $L_{JJK}$ ~/mHz    &         $-$2&.486(91)     &           $[B_{046}]_{10}$ ~/\textmu Hz    &          16&.92(68) \\
                     $l_{JK}$ ~/mHz    &         $-$1&.534(58)     &           $[B_{226}]_{10}$ ~/\textmu Hz    &          2&.129(122) \\
          $[B^-_{010}]_{1-1}$ ~/kHz    &         $-$885&.7(160)    &                $[A-(B+C)/2]_{11}$ ~/kHz    &         $-$209&.2(49) \\
          $[B^-_{100}]_{1-1}$ ~/MHz    &         $-$4&.156(78)     &               $[\Delta_{K}]_{11}$ ~/kHz    &          6&.362(61) \\
          $[A-(B+C)/2]_{1-1}$ ~/kHz    &          26&.08(141)      &                  $[\Phi_{K}]_{11}$ ~/Hz    &        $-$17&.702(210) \\
         $[\Delta_{K}]_{1-1}$ ~/kHz    &         1&.1225(199)      &                  $[(B+C)/2]_{11}$ ~/kHz    &          24&.29(131) \\
            $[\Phi_{K}]_{1-1}$ ~/Hz    &         $-$1&.380(59)     &              $[\Delta_{JK}]_{11}$ ~/kHz    &        $-$1&.3051(251) \\
            $[(B+C)/2]_{1-1}$ ~/kHz    &         $-$38&.80(89)     &                  $[\Delta_J]_{11}$ ~/Hz    &         $-$202&.7(80) \\
        $[\Delta_{JK}]_{1-1}$ ~/kHz    &        $-$1&.2087(213)    &                   $[\Phi_J]_{11}$ ~/mHz    &          441&.0(164) \\
            $[\Delta_J]_{1-1}$ ~/Hz    &          182&.4(54)       &                  $[L_{JJK}]_{11}$ ~/mHz    &          4&.529(170) \\
           $[\Phi_{JK}]_{1-1}$ ~/Hz    &          3&.210(65)       &             $[L_{J}]_{11}$ ~/\textmu Hz    &           $-$436&(35) \\
              $[\Phi_J]_{1-1}$ ~/Hz    &        $-$1&.6468(297)    &                  $[(B-C)/4]_{11}$ ~/kHz    &         $-$32&.49(40) \\
            $[L_{JJK}]_{1-1}$ ~/mHz    &         $-$1&.899(109)    &               $[\delta_{K}]_{11}$ ~/kHz    &          21&.08(30) \\
              $[L_{J}]_{1-1}$ ~/mHz    &          1&.487(63)       &                  $[\phi_{K}]_{11}$ ~/Hz    &         $-$70&.98(111) \\
          $[\delta_{K}]_{1-1}$ ~/Hz    &          426&.7(81)       &                    $[l_{K}]_{11}$ ~/mHz    &          68&.39(156) \\
            $[\phi_{K}]_{1-1}$ ~/Hz    &         $-$1&.497(115)    &                $[\delta_{J}]_{11}$ ~/Hz    &          $-$86&.9(40) \\
          $[\delta_{J}]_{1-1}$ ~/Hz    &          84&.56(261)      &                 $[\phi_{JK}]_{11}$ ~/Hz    &          2&.931(95) \\
           $[\phi_{J}]_{1-1}$ ~/mHz    &         $-$541&.3(116)    &                 $[\phi_{J}]_{11}$ ~/mHz    &          266&.7(91) \\
      $[l_{JK}]_{1-1}$ ~/\textmu Hz    &         $-$491&.6(293)    &             $[l_{J}]_{11}$ ~/\textmu Hz    &         $-$191&.0(215) \\
       $[l_{J}]_{1-1}$ ~/\textmu Hz    &          631&.4(227)      &                   $[B_{024}]_{11}$ ~/Hz    &          3&.543(98) \\
            $[B_{024}]_{1-1}$ ~/mHz    &         $-$879&.4(244     &                  $[B_{224}]_{11}$ ~/mHz    &          $-$2&.40(32) \\
     $[B_{044}]_{1-1}$ ~/\textmu Hz    &           $-$673&(47)     &                  $[B_{026}]_{11}$ ~/mHz    &          $-$5&.68(38) \\
                $[h_2]_{1-1}$ ~/mHz    &          641&.6(113)      &               $[l_3]_{11}$ ~/\textmu Hz    &          149&.2(212) \\
         $[l_2]_{1-1}$ ~/\textmu Hz    &         $-$566&.7(299)    &               $[l_4]_{11}$ ~/\textmu Hz    &          177&.0(150) \\
                $[h_3]_{1-1}$ ~/mHz    &          356&.5(56)       &                $[A-(B+C)/2]_{20}$ ~/kHz    &          285&.8(48) \\
         $[l_3]_{1-1}$ ~/\textmu Hz    &         $-$615&.8(249)    &               $[\Delta_{K}]_{20}$ ~/kHz    &         $-$1&.339(59) \\
         $[l_4]_{1-1}$ ~/\textmu Hz    &         $-$160&.3(59)     &                  $[\Phi_{K}]_{20}$ ~/Hz    &         $-$1&.901(133) \\
           $[B^-_{010}]_{10}$ ~/MHz    &         $-$25&.80(45)     &                  $[(B+C)/2]_{20}$ ~/kHz    &         $-$205&.0(43) \\
           $[B^-_{210}]_{10}$ ~/kHz    &          44&.06(61)       &              $[\Delta_{JK}]_{20}$ ~/kHz    &          2&.824(88) \\
           $[A-(B+C)/2]_{10}$ ~/MHz    &        $-$1&.9353(232)    &                 $[\Phi_{KJ}]_{20}$ ~/Hz    &          2&.226(101) \\
          $[\Delta_{K}]_{10}$ ~/kHz    &         $-$4&.351(84)     &                 $[\Delta_J]_{20}$ ~/kHz    &         $-$1&.251(32) \\
             $[\Phi_{K}]_{10}$ ~/Hz    &          $-$1&.13(38)     &                  $[(B-C)/4]_{20}$ ~/kHz    &          507&.8(70) \\
        $[M_{K}]_{10}$ ~/\textmu Hz    &          250&.2(207)      &                $[\delta_{K}]_{20}$ ~/Hz    &            251&(32) \\
             $[B_{100}]_{10}$ ~/MHz    &          137&.0(40)       &                $[\delta_{J}]_{20}$ ~/Hz    &         $-$46&.42(224) \\
             $[(B+C)/2]_{10}$ ~/kHz    &          373&.4(57)       &                $[\phi_{JK}]_{20}$ ~/mHz    &          597&.0(264) \\
         $[\Delta_{JK}]_{10}$ ~/kHz    &          4&.526(87)       &                 $[\phi_{J}]_{20}$ ~/mHz    &         $-$239&.0(113) \\
            $[\Phi_{KJ}]_{10}$ ~/Hz    &           6&.77(63)       &                       $[d_2]_{20}$ ~/Hz    &          761&.2(193) \\
      $[M_{KKJ}]_{10}$ ~/\textmu Hz    &           $-$518&(39)     &                      $[h_3]_{20}$ ~/mHz    &          197&.2(76) \\      
            $[\Phi_{JK}]_{10}$ ~/Hz    &          $-$8&.38(34)     \\ 
                                                                                                                   \\

    \hline

no. of transitions fit      &  1621&          \\
no. of lines fit            &  1298&           \\
standard deviation$^b$      &     1&.36     \\
    \hline   
  \end{tabular}\\[2pt]
}
\end{center}

$^a$\footnotesize{Numbers in parentheses are one standard deviation in
units of the least significant figures.}
$^b$\footnotesize{Weighted unitless value for the entire fit.}
\end{table*}


\begin{table}
\begin{center}

  \caption{Spectroscopic parameters$^a$ of the vibrationally
excited state $\upsilon_{17}=1$ of CH$_3$C(O)CH$_3$.}
  \label{A:spectroscopic-parameters-12C-v24}
{\tiny
  \begin{tabular}{lr@{}l}
  \hline

    Parameter & \multicolumn{2}{c}{$\upsilon_{17}=1$}    \\
     \hline
     \\                          

                             $\rho$    &         0&.062788(80) \\
               $\beta$ ~/$^{\circ}$    &         27&.100(45) \\
                          $A$ ~/MHz    &      10191&.123(98) \\
                          $B$ ~/MHz    &       8480&.859(81) \\
                          $C$ ~/MHz    &     4910&.06430(229) \\
                 $\Delta_{J}$ ~/kHz    &          7&.812(109) \\
                $\Delta_{JK}$ ~/kHz    &           2&.84(118) \\
                 $\Delta_{K}$ ~/kHz    &         $-$37&.67(191) \\
                 $\delta_{J}$ ~/kHz    &          3&.535(54) \\
                 $\delta_{K}$ ~/kHz    &           2&.57(46) \\
                   $\Phi_{J}$ ~//Hz    &         10&.314(283) \\
                  $\Phi_{JK}$ ~//Hz    &         $-$41&.17(111) \\
                  $\Phi_{KJ}$ ~//Hz    &          $-$6&.42(31) \\
                   $\Phi_{K}$ ~//Hz    &         $-$42&.93(178) \\
                    $\phi_{J}$ ~/Hz    &          5&.153(141) \\
             $\epsilon_{1-1}$ ~/MHz    &         $-$30&.21(74) \\
              $\epsilon_{10}$ ~/GHz    &        12&.9700(287) \\
              $\epsilon_{11}$ ~/MHz    &         102&.91(295) \\
              $\epsilon_{20}$ ~/MHz    &          458&.9(32) \\
              $\epsilon_{30}$ ~/MHz    &          21&.24(32) \\
          $[A-(B+C)/2]_{1-1}$ ~/kHz    &         $-$302&.8(68) \\
            $[(B+C)/2]_{1-1}$ ~/kHz    &          36&.29(68) \\
        $[\Delta_{JK}]_{1-1}$ ~/kHz    &         1&.0376(295) \\
            $[\Delta_J]_{1-1}$ ~/Hz    &         $-$395&.6(117) \\
          $[\delta_{K}]_{1-1}$ ~/Hz    &          341&.7(136) \\
          $[\delta_{J}]_{1-1}$ ~/Hz    &         $-$193&.4(59) \\
           $[B^-_{010}]_{10}$ ~/MHz    &          43&.83(282) \\
           $[A-(B+C)/2]_{10}$ ~/kHz    &           $-$461&(97) \\
          $[\Delta_{K}]_{10}$ ~/kHz    &         $-$17&.97(54) \\
             $[(B+C)/2]_{10}$ ~/kHz    &          812&.4(211) \\
             $[B_{210}]_{10}$ ~/kHz    &           8&.41(40) \\
         $[\Delta_{JK}]_{10}$ ~/kHz    &         13&.051(258) \\
             $[\Delta_J]_{10}$ ~/Hz    &         $-$882&.8(224) \\
            $[\Phi_{JK}]_{10}$ ~/Hz    &           6&.45(32) \\
             $[B_{011}]_{10}$ ~/kHz    &           $-$437&(69) \\
             $[B_{101}]_{10}$ ~/MHz    &         $-$5&.693(107) \\
             $[B_{211}]_{10}$ ~/kHz    &         $-$2&.599(194) \\
             $[(B-C)/4]_{10}$ ~/kHz    &          339&.0(98) \\
             $[\phi_{K}]_{10}$ ~/Hz    &          24&.97(86) \\
            $[\phi_{J}]_{10}$ ~/mHz    &         $-$299&.8(219) \\
                  $[d_2]_{10}$ ~/Hz    &          468&.6(114) \\
              $[B_{024}]_{10}$ ~/Hz    &         $-$3&.529(146) \\
                 $[h_2]_{10}$ ~/mHz    &           $-$482&(35) \\
                 $[h_3]_{10}$ ~/mHz    &         $-$182&.5(126) \\
             $[(B+C)/2]_{11}$ ~/kHz    &          108&.9(66) \\
             $[\Delta_J]_{11}$ ~/Hz    &           7&.58(48) \\
             $[(B-C)/4]_{11}$ ~/kHz    &           61&.4(31) \\
           $[\delta_{K}]_{11}$ ~/Hz    &          314&.4(182) \\
           $[A-(B+C)/2]_{20}$ ~/kHz    &         $-$298&.2(261) \\
          $[\Delta_{K}]_{20}$ ~/kHz    &          2&.654(88) \\
             $[(B+C)/2]_{20}$ ~/kHz    &          263&.3(59) \\
           $[\delta_{K}]_{20}$ ~/Hz    &          966&.6(249) \\ 
                                                                                                                   \\

    \hline

no. of transitions fit      &   939&           \\
no. of lines fit            &   719&           \\
standard deviation$^b$      &     1&.62        \\
    \hline   
  \end{tabular}\\[2pt]
}
\end{center}

$^a$\footnotesize{Numbers in parentheses are one standard deviation
in units of the least significant figures.} 
$^b$\footnotesize{Weighted unitless value for the entire fit.}
\end{table}


\section{Complementary Figures}
\label{a:figures}
Figures~\ref{f:spec_ch3coch3_ve0}--\ref{f:spec_ch3coch3_13c2_ve0} show the
transitions of CH$_3$C(O)CH$_3$ $\varv=0$, $\varv_{12}=1$, and $\varv_{17}=1$,
as well as $^{13}$CH$_3$C(O)CH$_3$ $\varv=0$ and CH$_3$ $^{13}$C(O)CH$_3$ 
$\varv=0$ that are covered by the EMoCA survey and contribute significantly to 
the signal detected toward Sgr~B2(N2). 

\newpage
~\newpage

\begin{figure*}
\centerline{\resizebox{0.83\hsize}{!}{\includegraphics[angle=0]{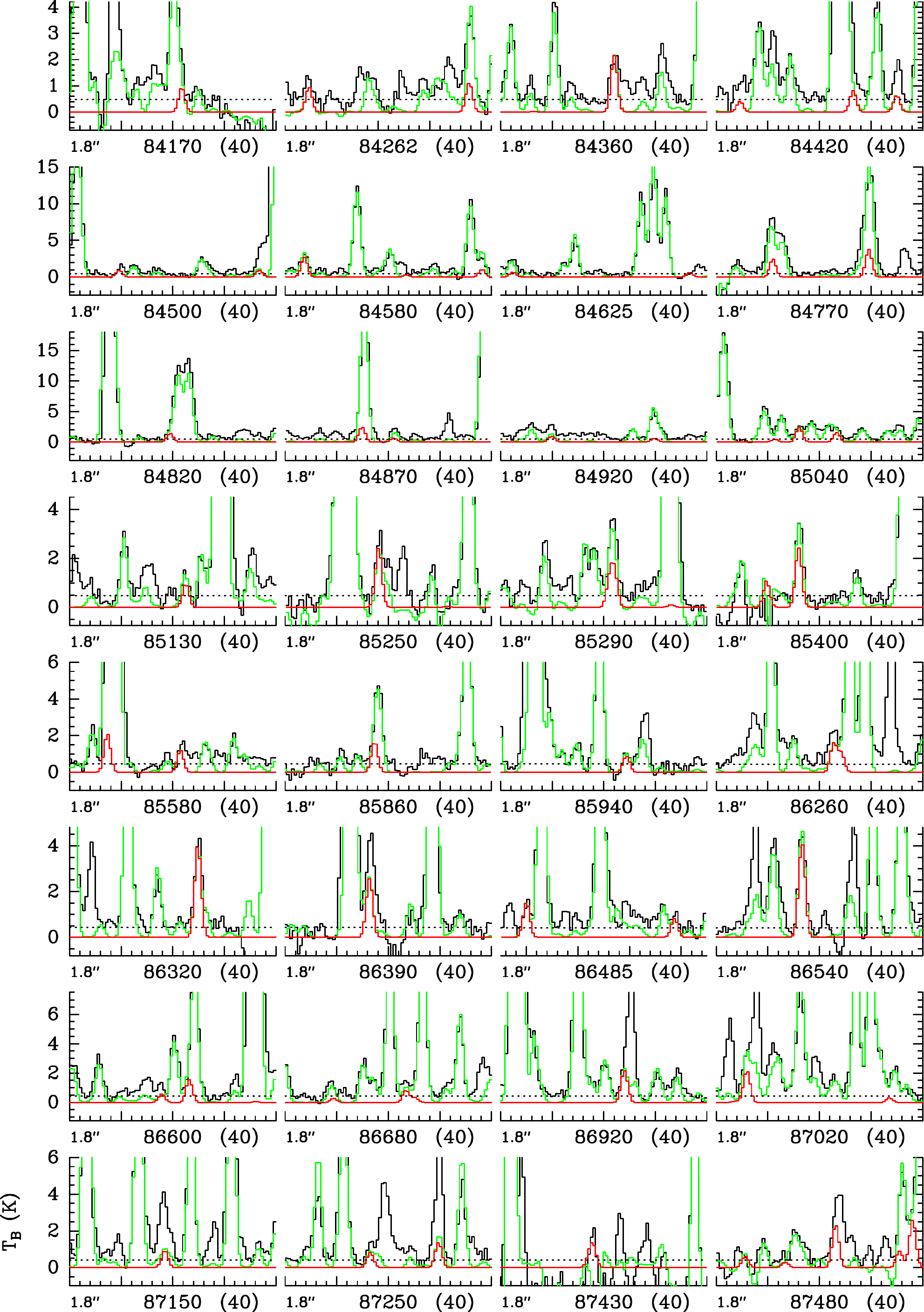}}}
\caption{
Transitions of CH$_3$C(O)CH$_3$ $\varv = 0$ covered by the EMoCA survey.
The best-fit LTE synthetic spectrum of CH$_3$C(O)CH$_3$ $\varv = 0$ 
is displayed in red and overlaid on the observed spectrum of Sgr~B2(N2) shown 
in black. The green synthetic spectrum contains the contributions of all 
molecules identified in our survey so far, including the species shown in red. 
The central frequency and width of the frequency range (in parentheses) are 
indicated in MHz below each panel. The angular resolution (HPBW) is also 
indicated. The y-axis is labeled in brightness temperature units. The 
dotted line indicates the $3\sigma$ noise level.
}
\label{f:spec_ch3coch3_ve0}
\end{figure*}

\clearpage
\begin{figure*}
\addtocounter{figure}{-1}
\centerline{\resizebox{0.83\hsize}{!}{\includegraphics[angle=0]{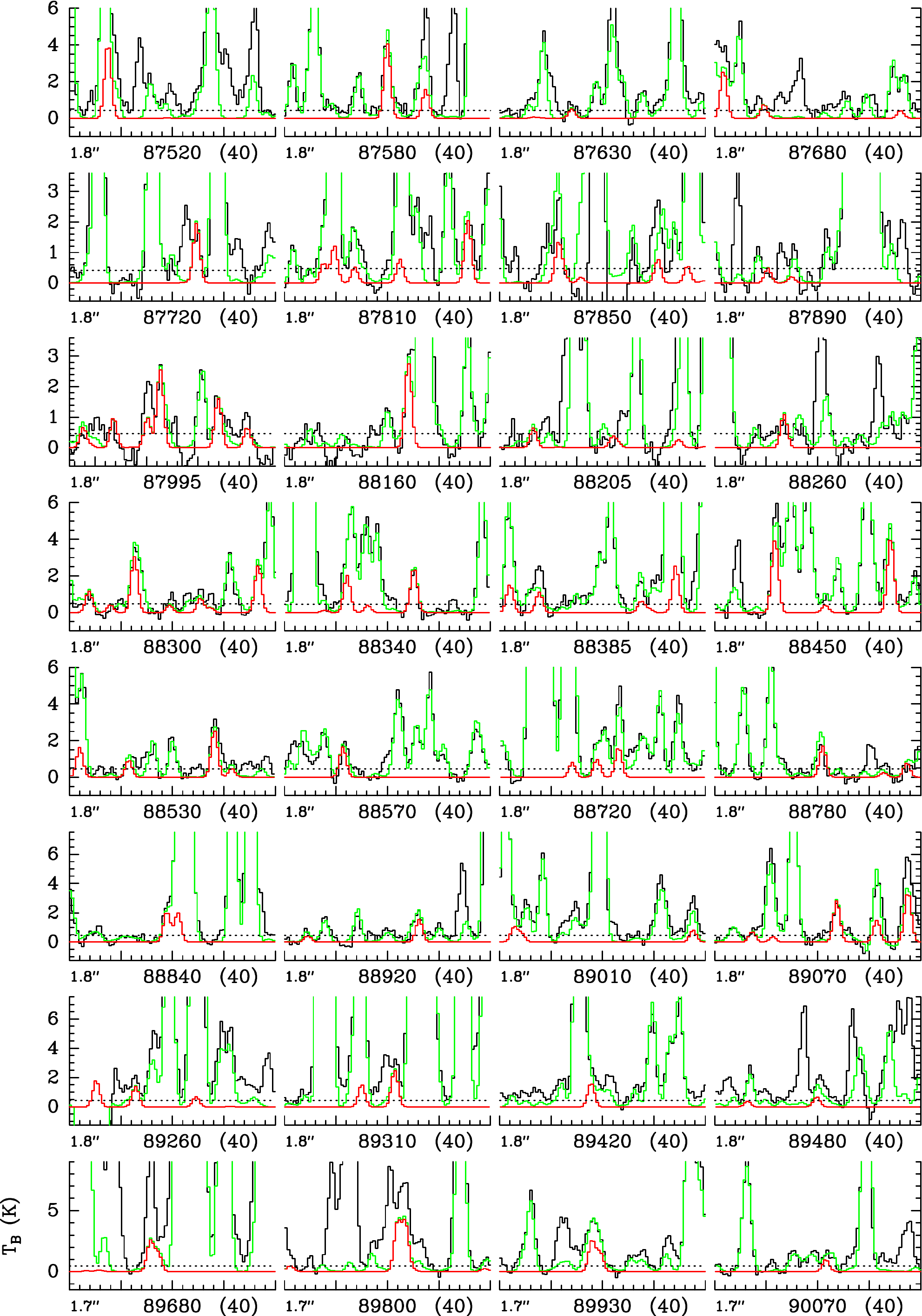}}}
\caption{continued.}
\end{figure*}

\clearpage
\begin{figure*}
\addtocounter{figure}{-1}
\centerline{\resizebox{0.83\hsize}{!}{\includegraphics[angle=0]{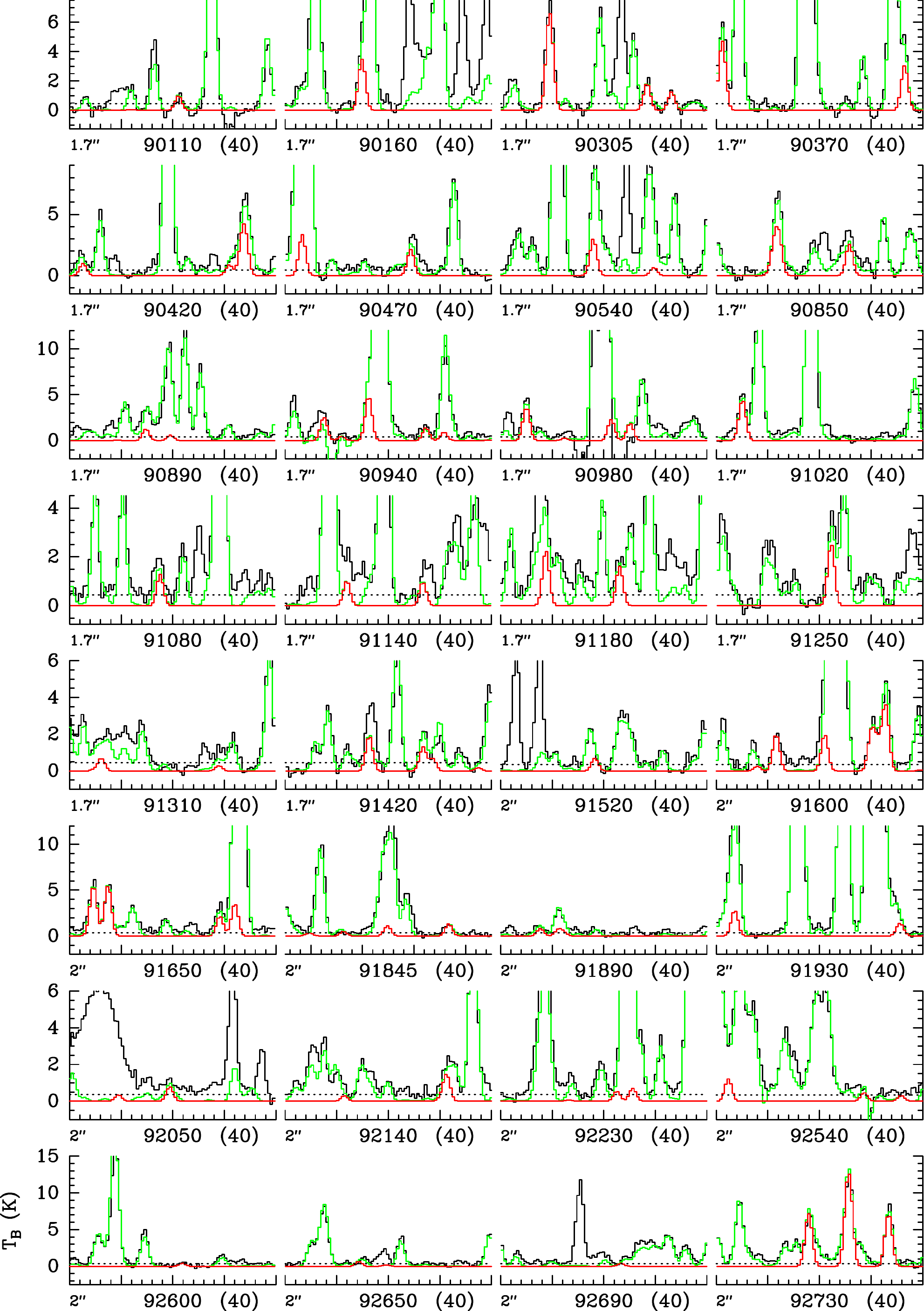}}}
\caption{continued.}
\end{figure*}

\clearpage
\begin{figure*}
\addtocounter{figure}{-1}
\centerline{\resizebox{0.83\hsize}{!}{\includegraphics[angle=0]{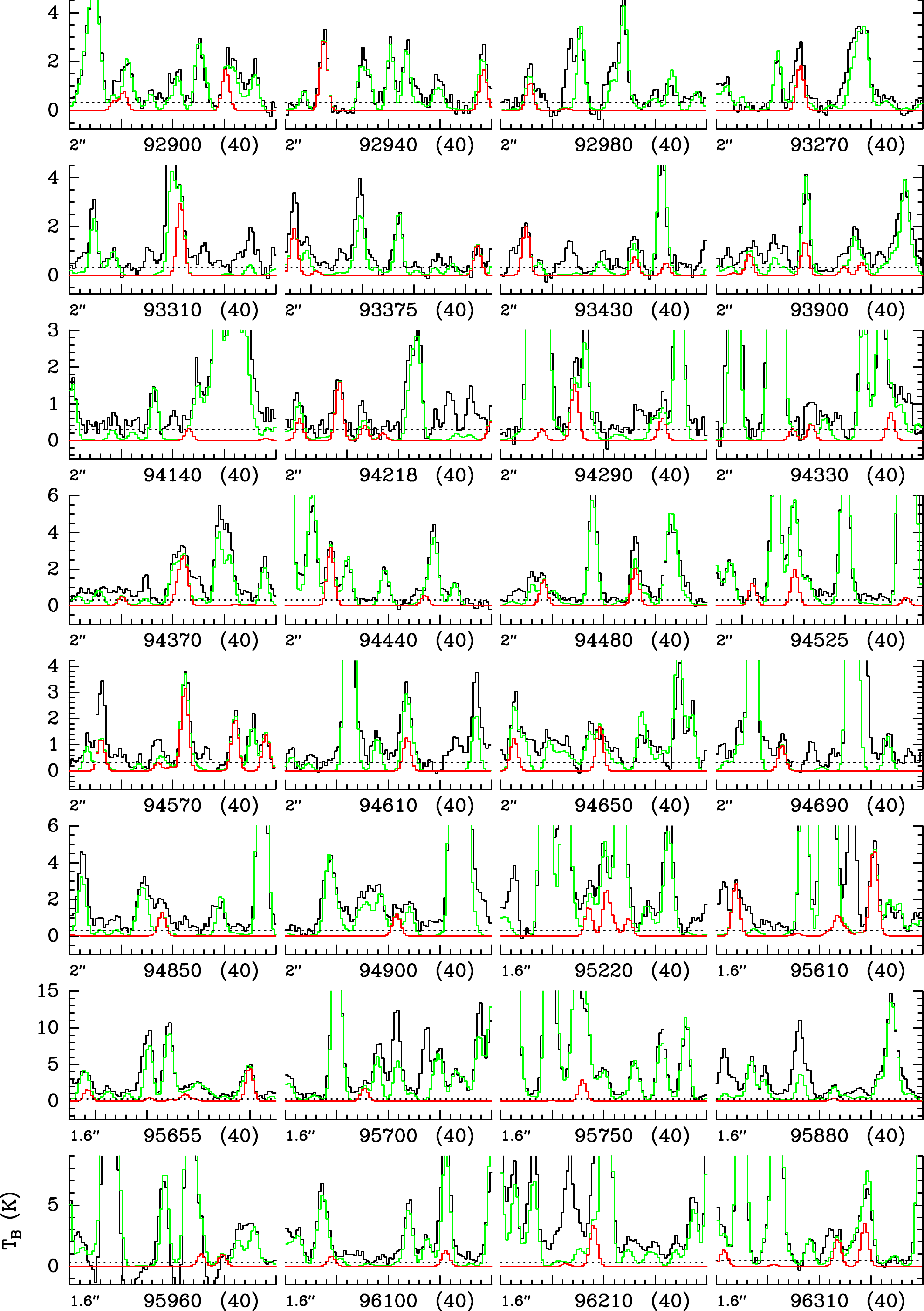}}}
\caption{continued.}
\end{figure*}

\clearpage
\begin{figure*}
\addtocounter{figure}{-1}
\centerline{\resizebox{0.83\hsize}{!}{\includegraphics[angle=0]{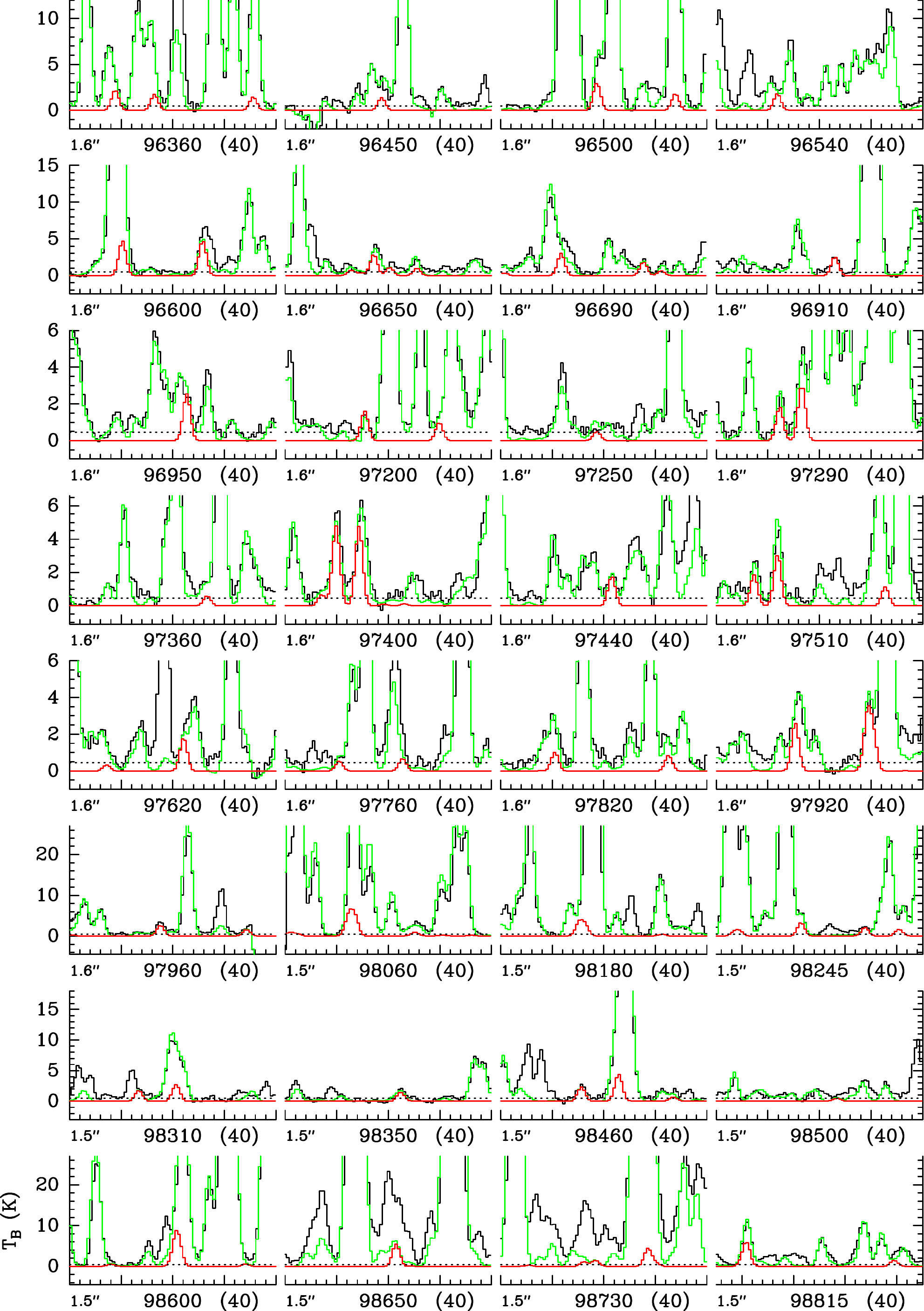}}}
\caption{continued.}
\end{figure*}

\clearpage
\begin{figure*}
\addtocounter{figure}{-1}
\centerline{\resizebox{0.83\hsize}{!}{\includegraphics[angle=0]{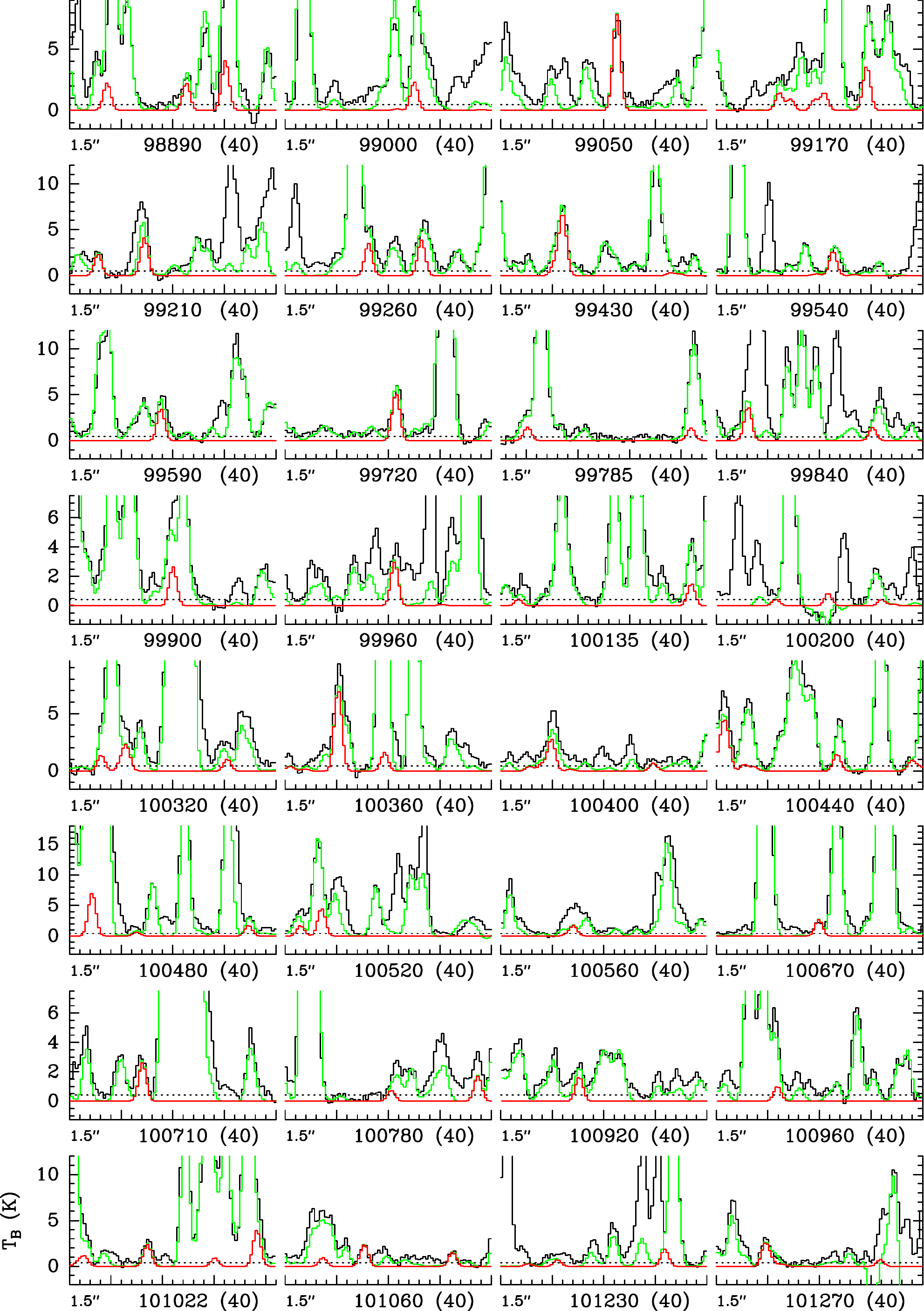}}}
\caption{continued.}
\end{figure*}

\clearpage
\begin{figure*}
\addtocounter{figure}{-1}
\centerline{\resizebox{0.83\hsize}{!}{\includegraphics[angle=0]{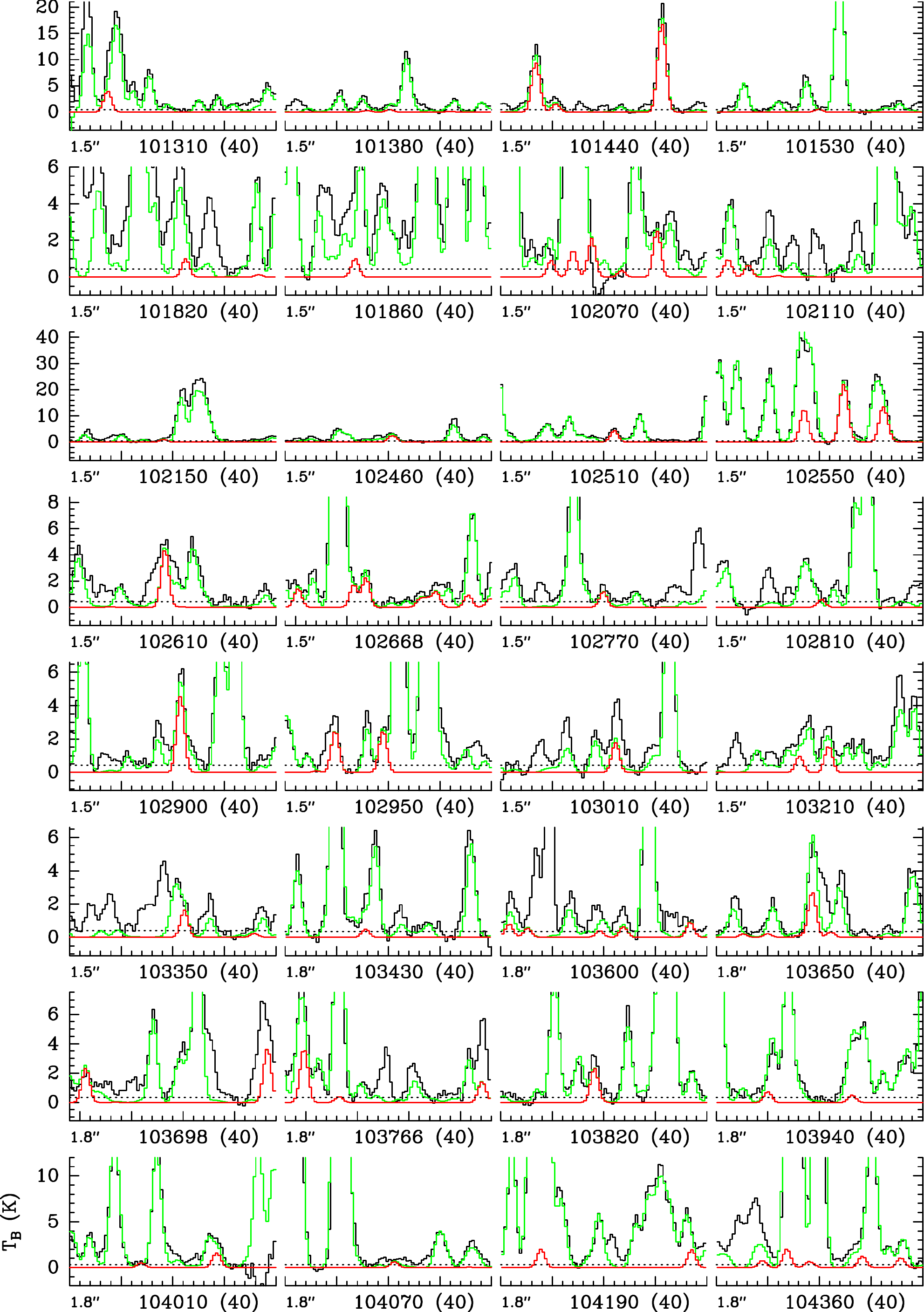}}}
\caption{continued.}
\end{figure*}

\clearpage
\begin{figure*}
\addtocounter{figure}{-1}
\centerline{\resizebox{0.83\hsize}{!}{\includegraphics[angle=0]{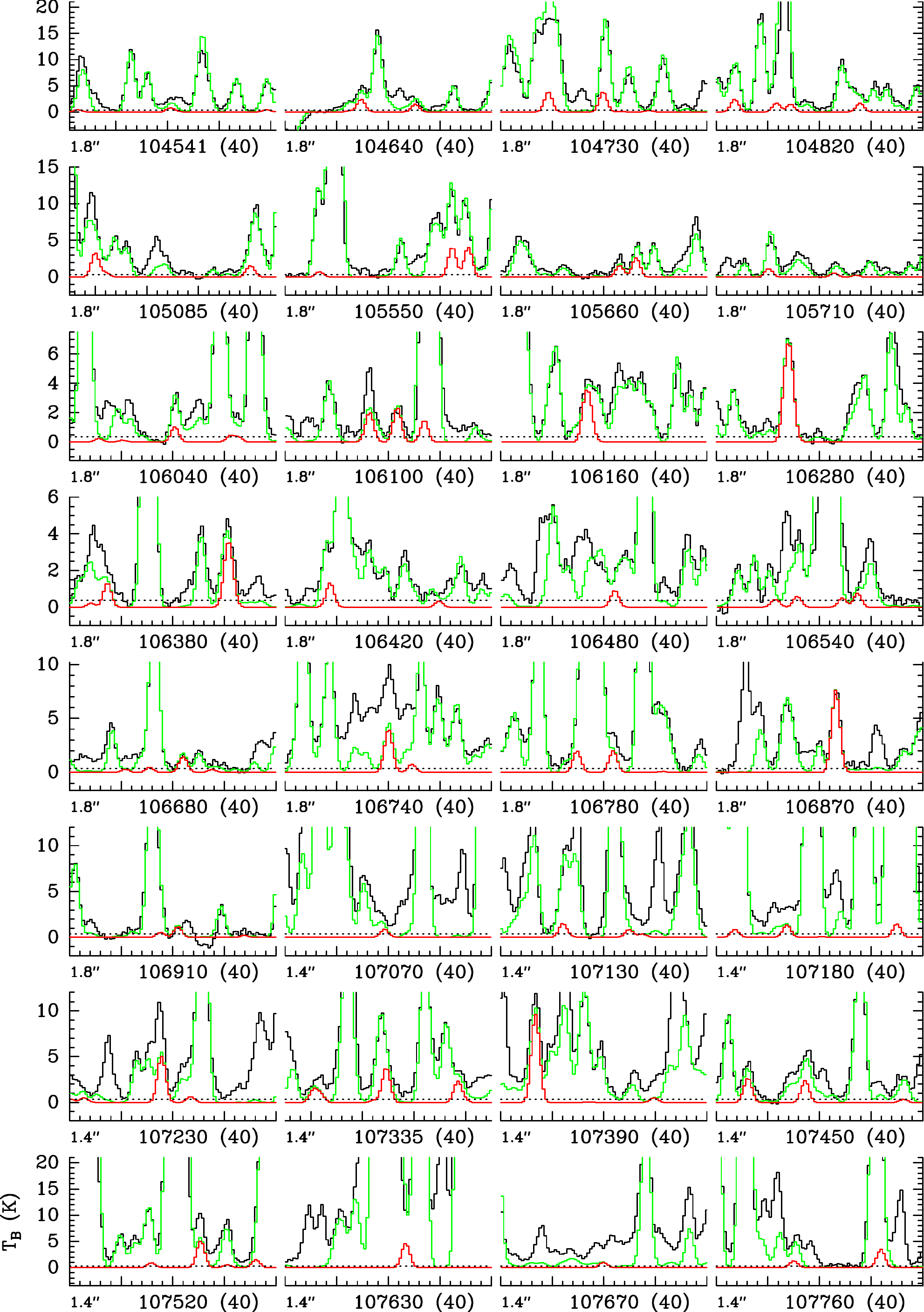}}}
\caption{continued.}
\end{figure*}

\clearpage
\begin{figure*}
\addtocounter{figure}{-1}
\centerline{\resizebox{0.83\hsize}{!}{\includegraphics[angle=0]{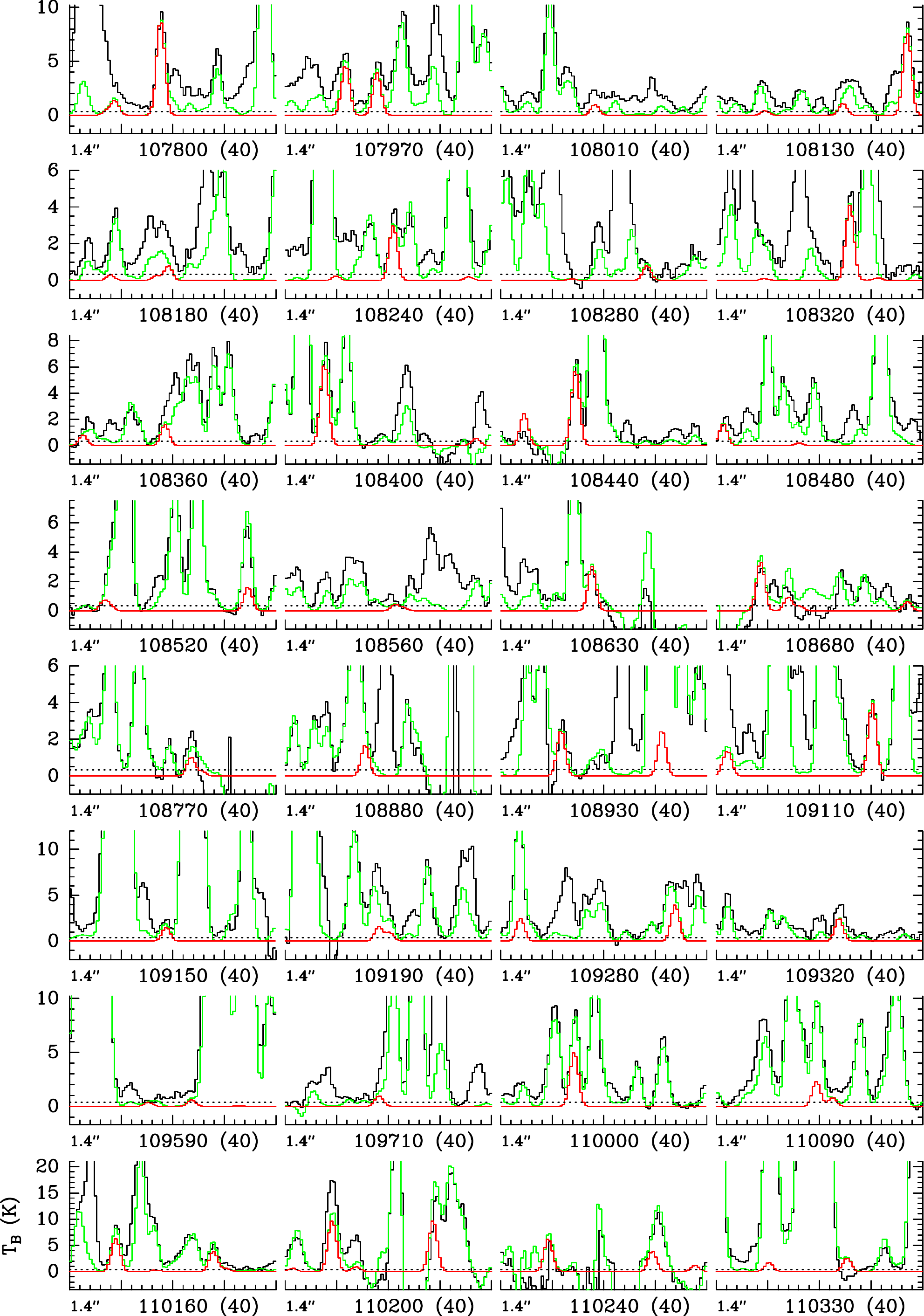}}}
\caption{continued.}
\end{figure*}

\clearpage
\begin{figure*}
\addtocounter{figure}{-1}
\centerline{\resizebox{0.83\hsize}{!}{\includegraphics[angle=0]{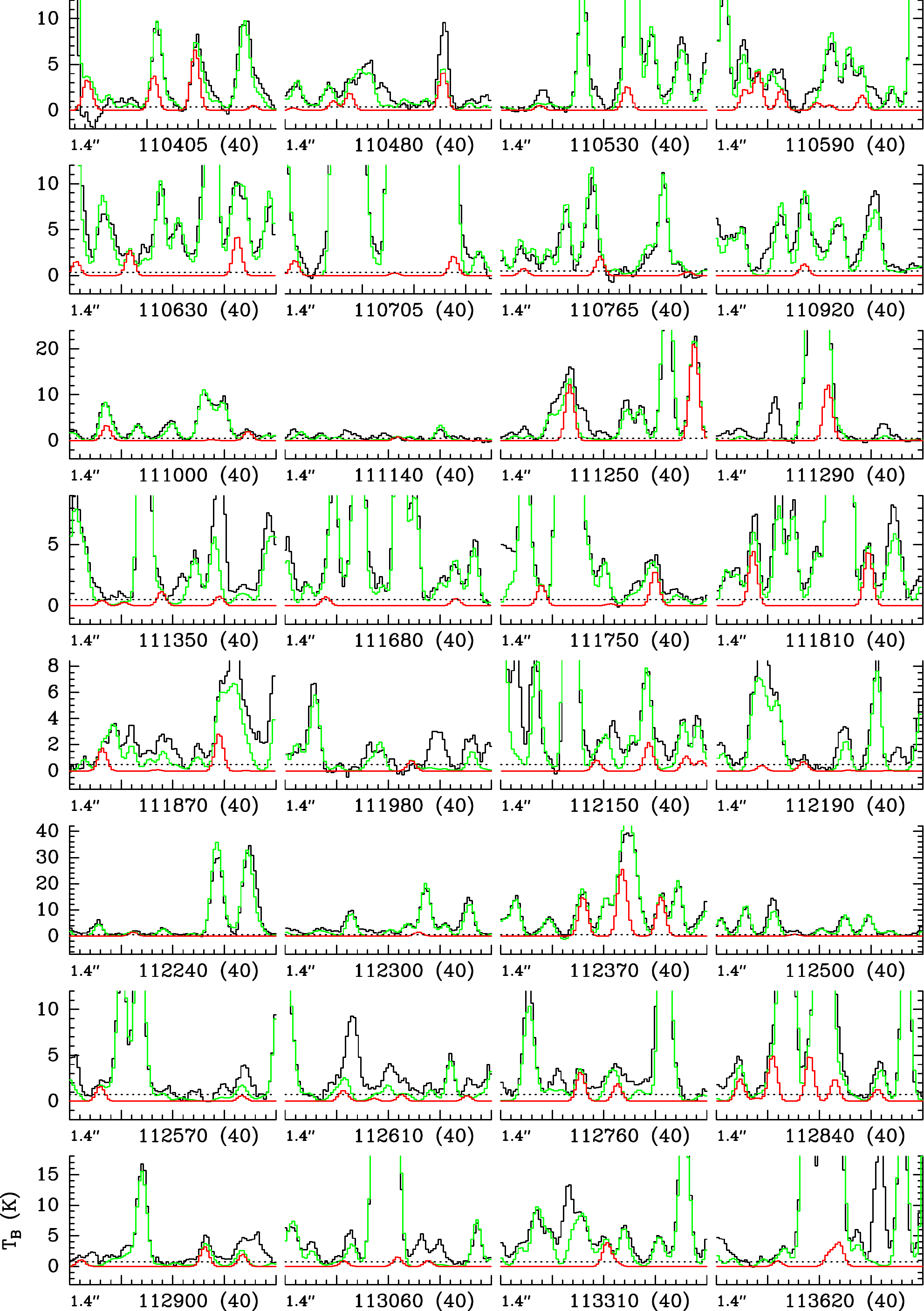}}}
\caption{continued.}
\end{figure*}

\clearpage
\begin{figure*}
\addtocounter{figure}{-1}
\centerline{\resizebox{0.83\hsize}{!}{\includegraphics[angle=0]{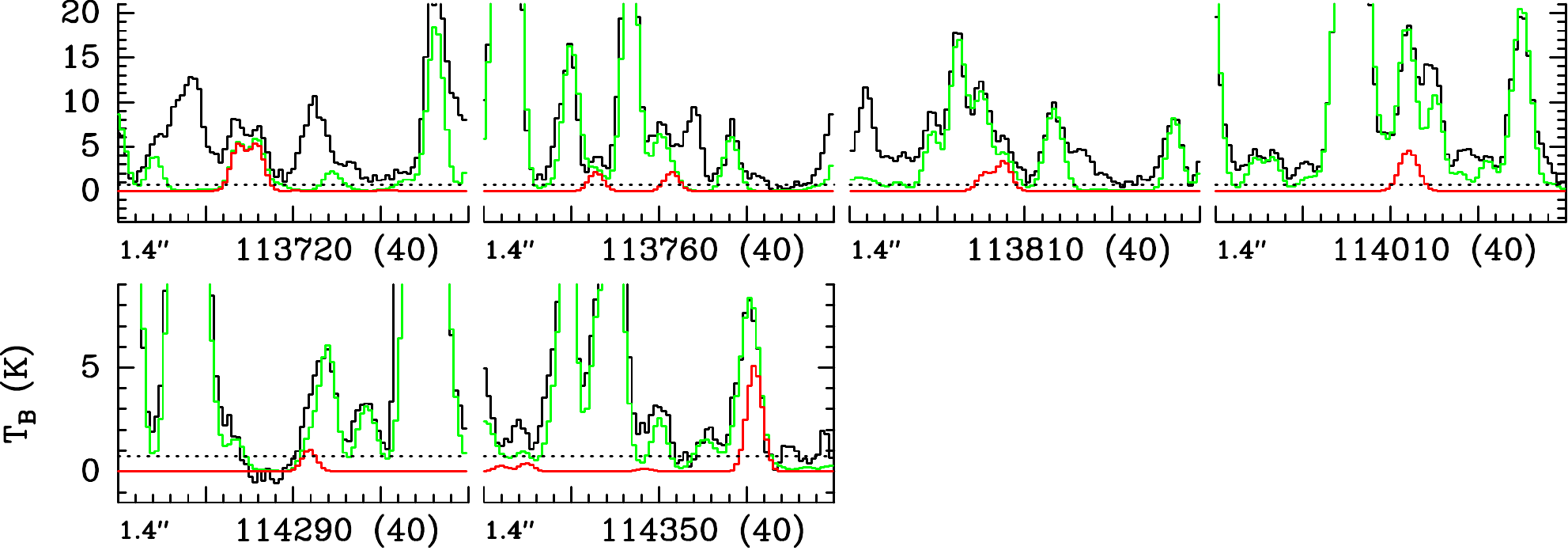}}}
\caption{continued.}
\end{figure*}

\clearpage
\begin{figure*}
\centerline{\resizebox{0.82\hsize}{!}{\includegraphics[angle=0]{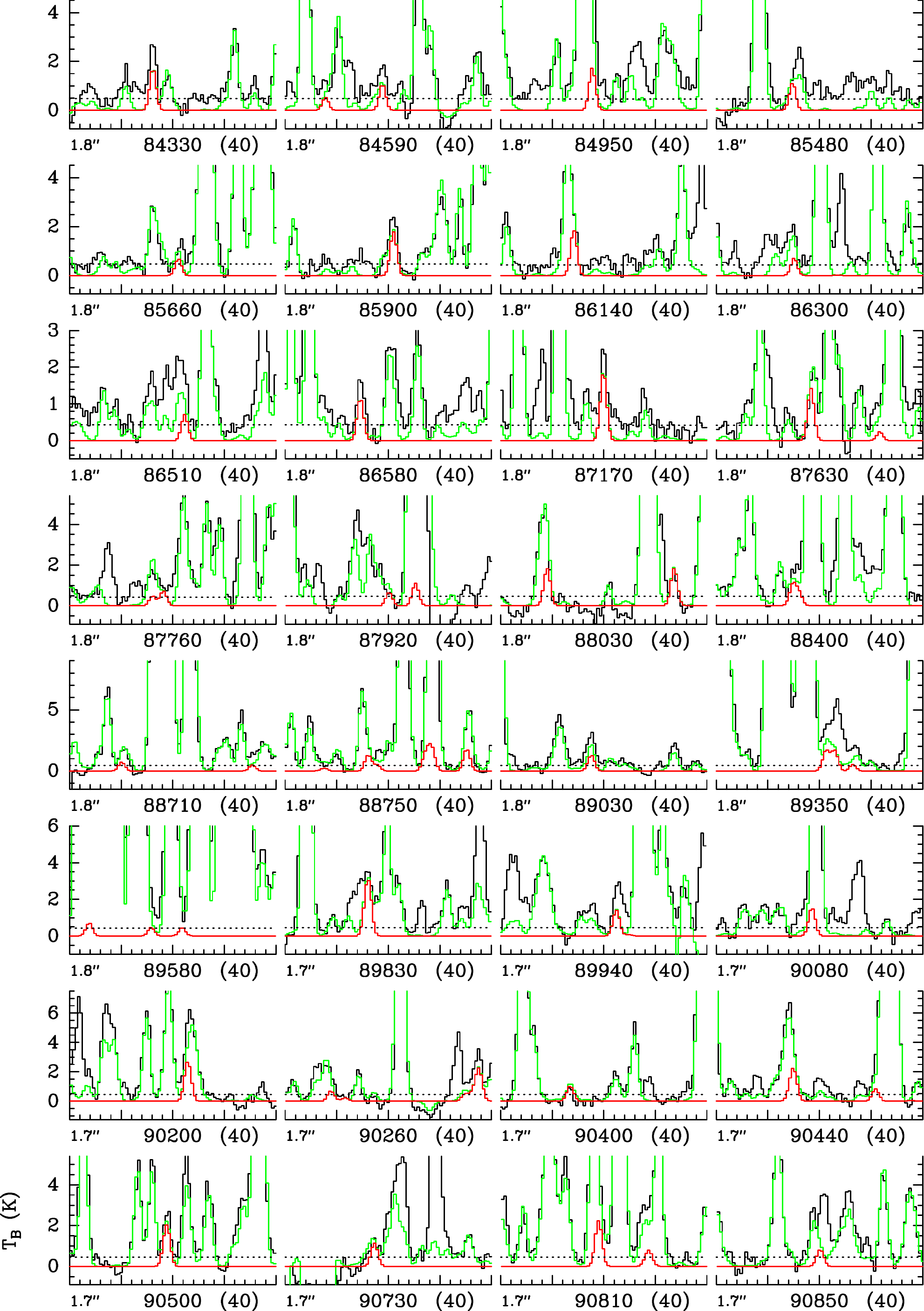}}}
\caption{Same as Fig.~\ref{f:spec_ch3coch3_ve0}, but for CH$_3$C(O)CH$_3$ 
$\varv_{12} = 1$.}
\label{f:spec_ch3coch3_v12e1}
\end{figure*}

\clearpage
\begin{figure*}
\addtocounter{figure}{-1}
\centerline{\resizebox{0.83\hsize}{!}{\includegraphics[angle=0]{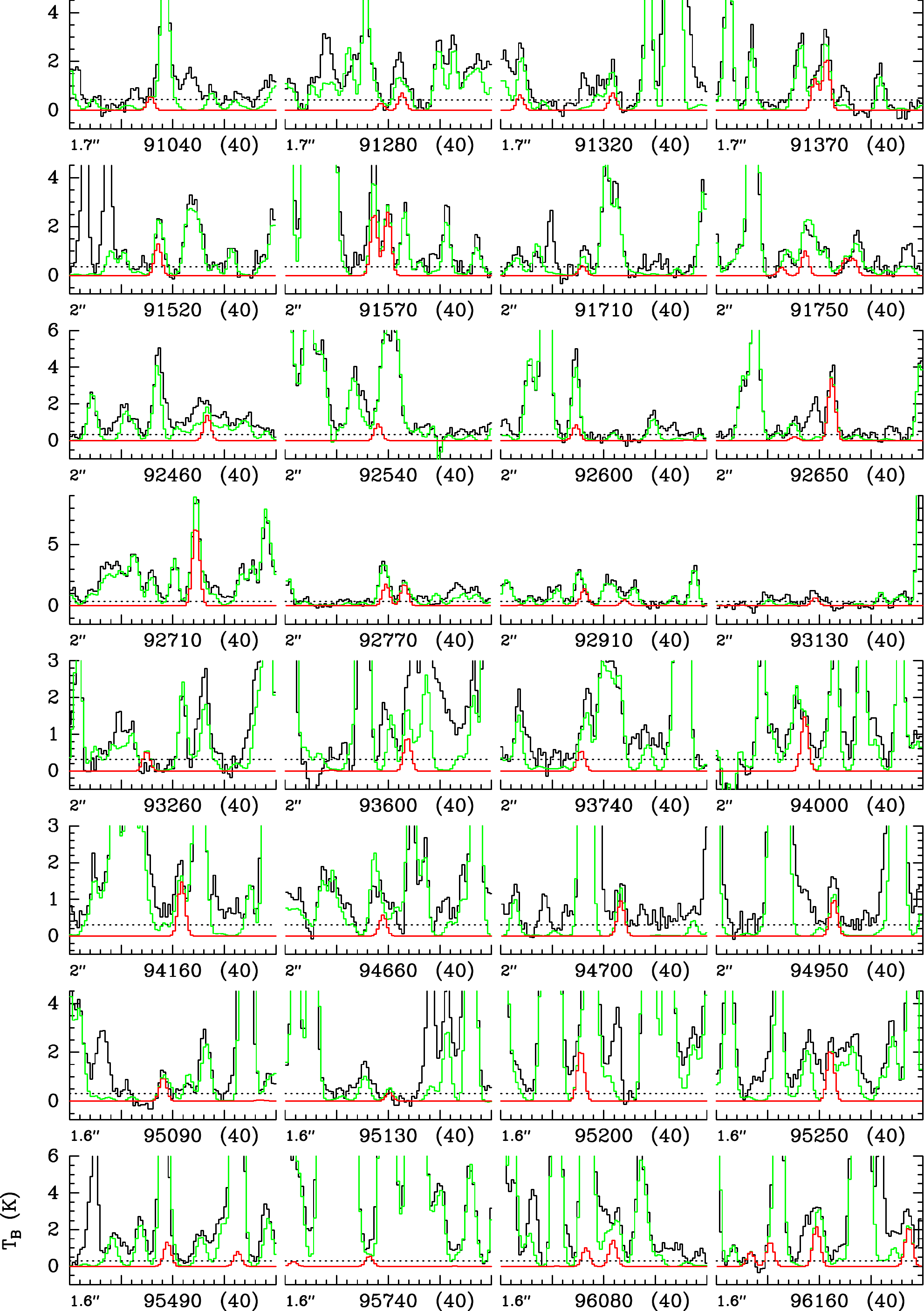}}}
\caption{continued.}
\end{figure*}

\clearpage
\begin{figure*}
\addtocounter{figure}{-1}
\centerline{\resizebox{0.83\hsize}{!}{\includegraphics[angle=0]{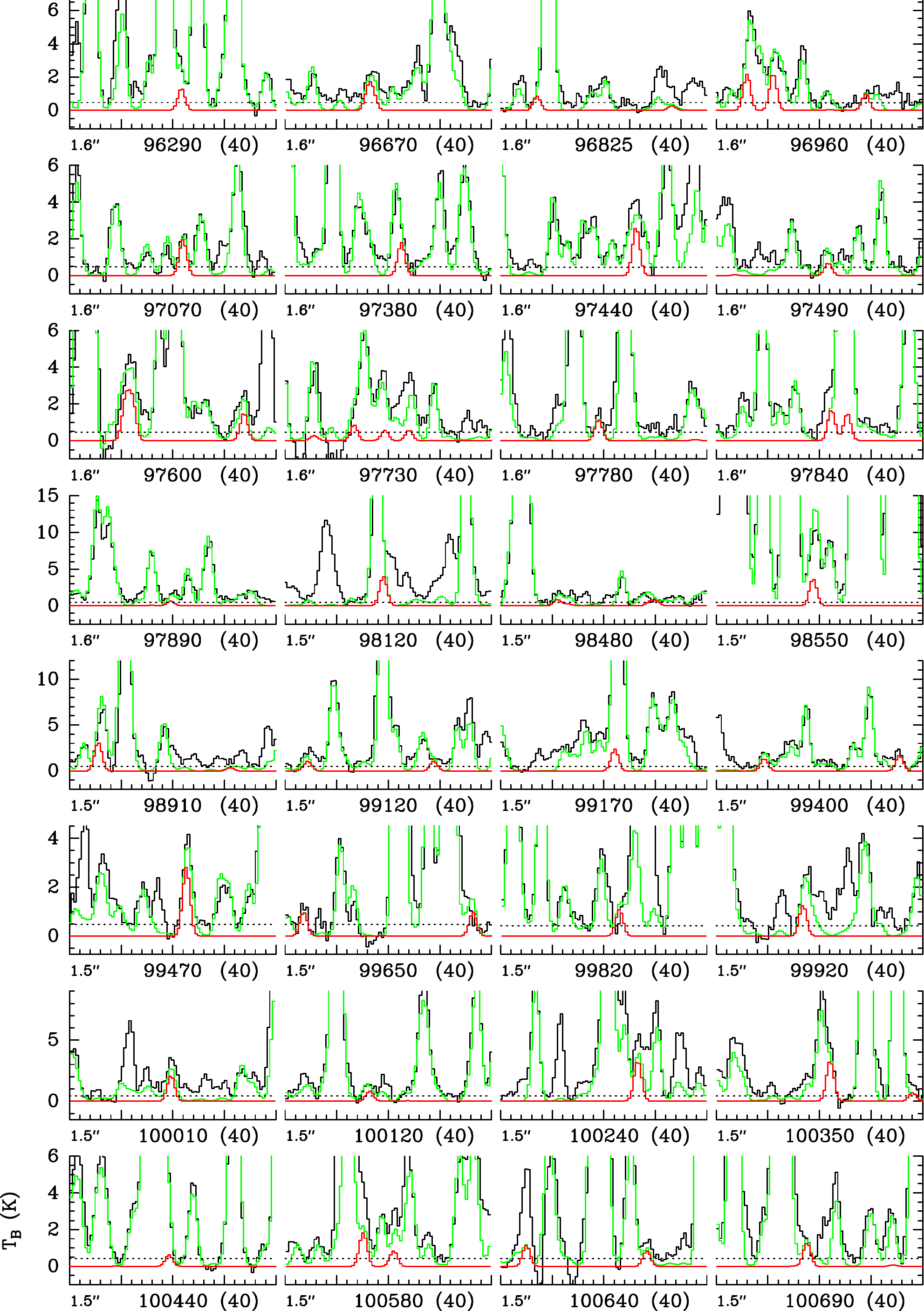}}}
\caption{continued.}
\end{figure*}

\clearpage
\begin{figure*}
\addtocounter{figure}{-1}
\centerline{\resizebox{0.83\hsize}{!}{\includegraphics[angle=0]{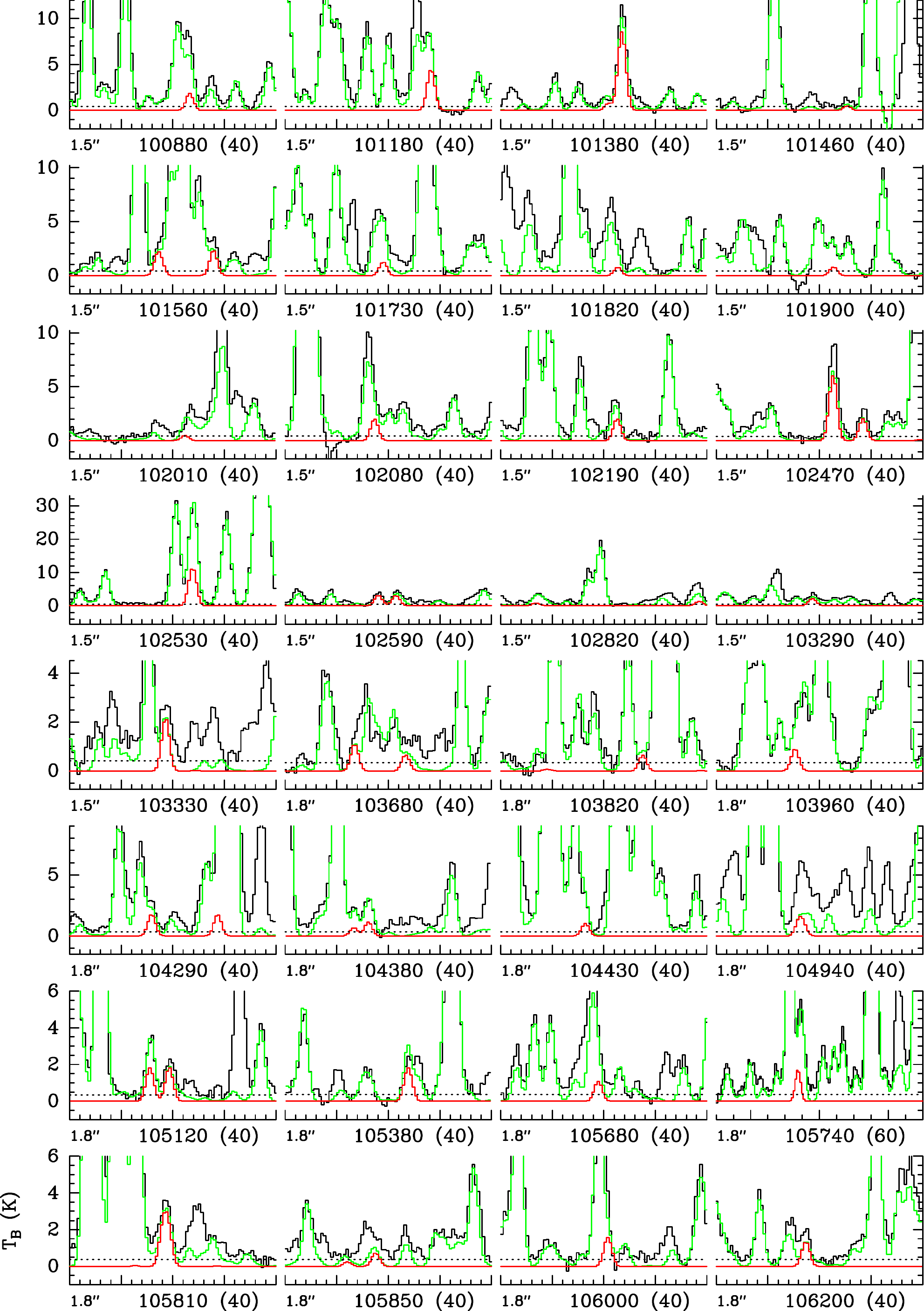}}}
\caption{continued.}
\end{figure*}

\clearpage
\begin{figure*}
\addtocounter{figure}{-1}
\centerline{\resizebox{0.83\hsize}{!}{\includegraphics[angle=0]{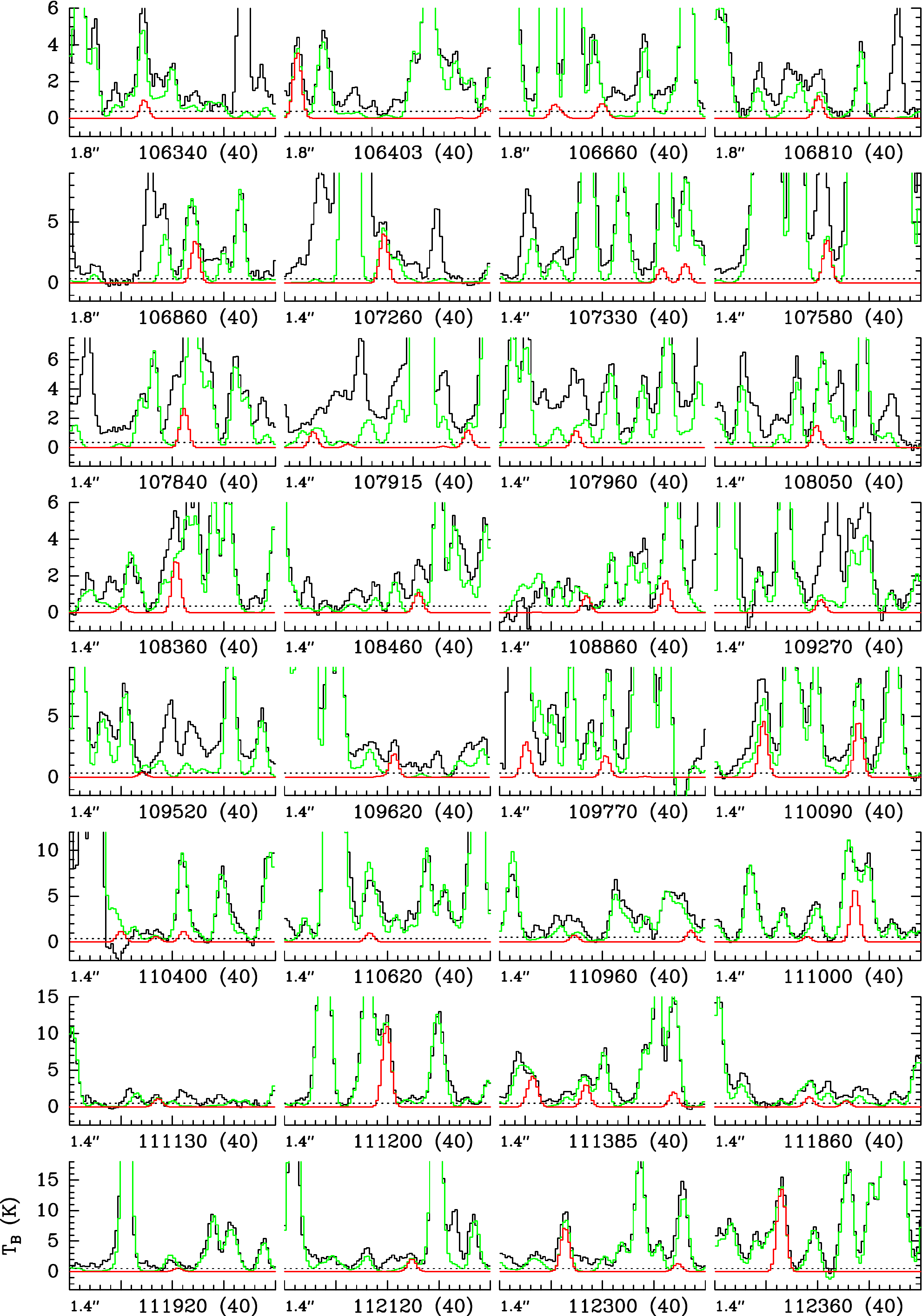}}}
\caption{continued.}
\end{figure*}

\clearpage
\begin{figure*}
\addtocounter{figure}{-1}
\centerline{\resizebox{0.83\hsize}{!}{\includegraphics[angle=0]{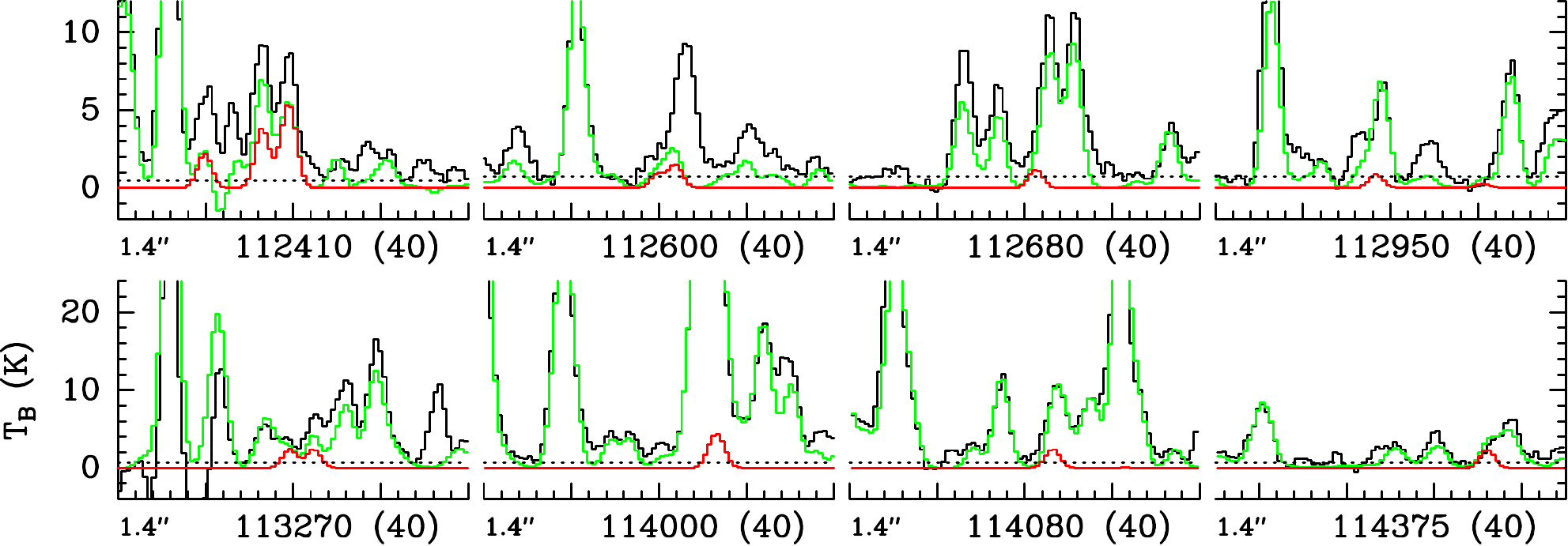}}}
\caption{continued.}
\end{figure*}

\clearpage
\begin{figure*}
\centerline{\resizebox{0.82\hsize}{!}{\includegraphics[angle=0]{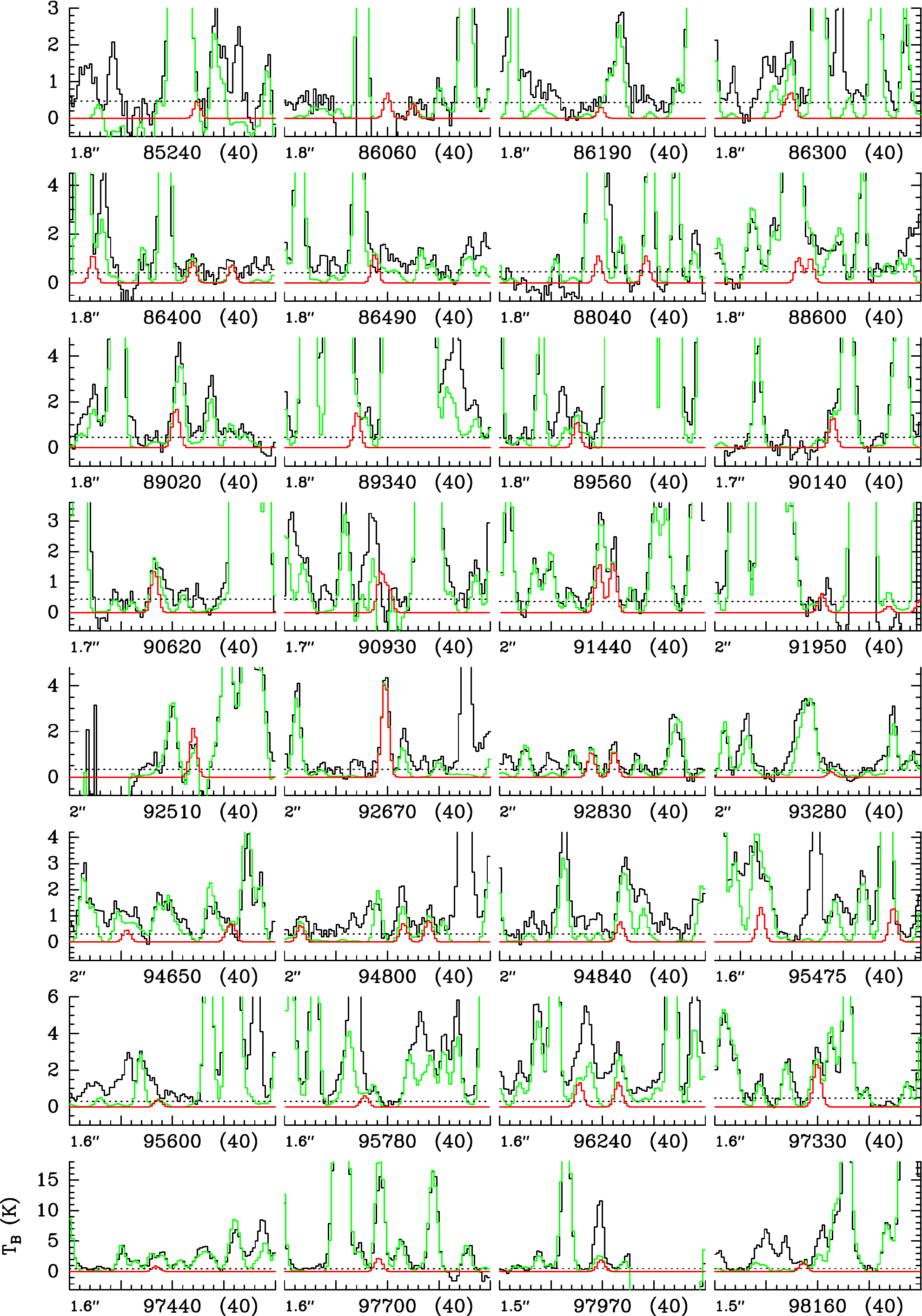}}}
\caption{Same as Fig.~\ref{f:spec_ch3coch3_ve0}, but for CH$_3$C(O)CH$_3$ 
$\varv_{17} = 1$.}
\label{f:spec_ch3coch3_v24e1}
\end{figure*}

\clearpage
\begin{figure*}
\addtocounter{figure}{-1}
\centerline{\resizebox{0.83\hsize}{!}{\includegraphics[angle=0]{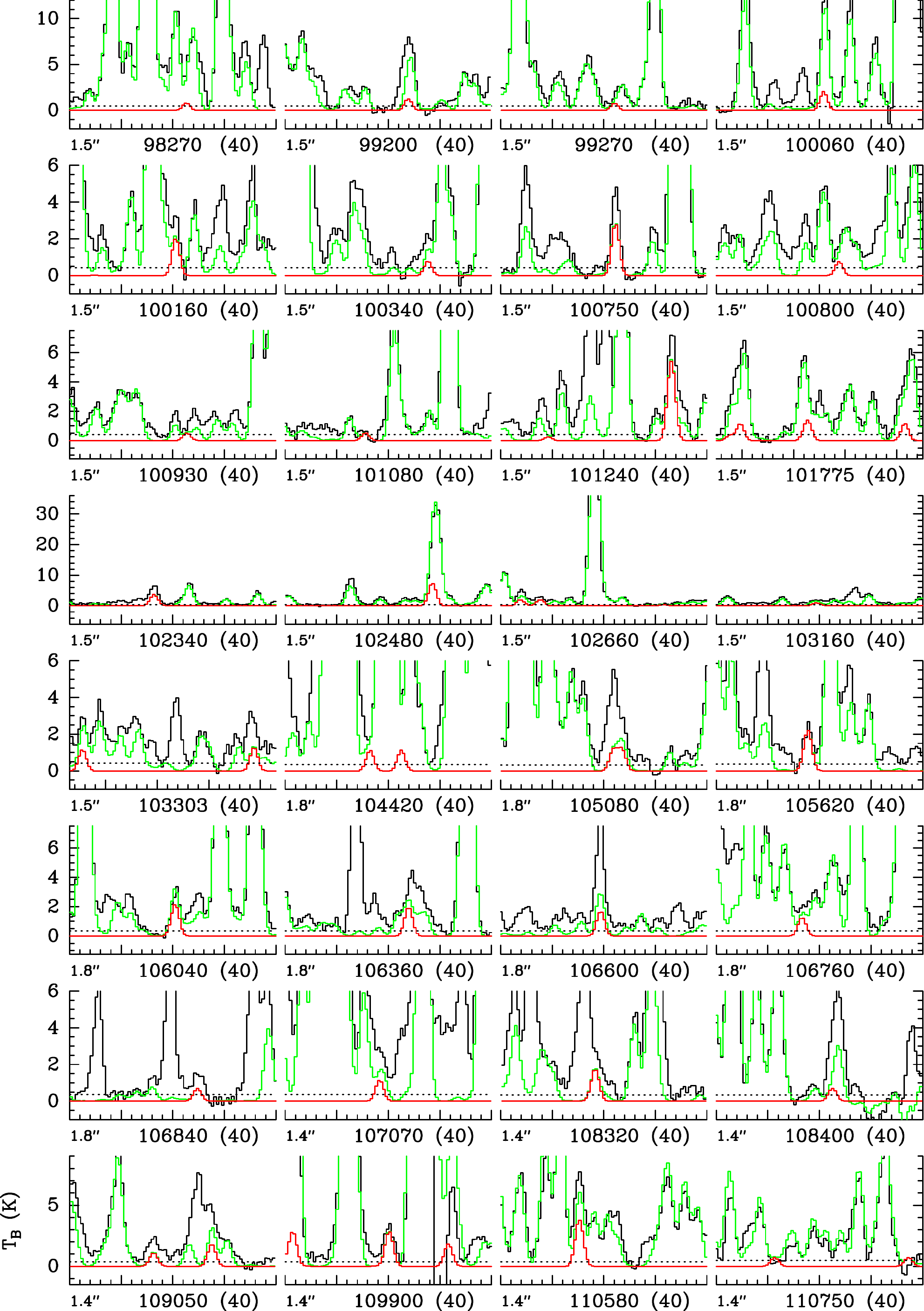}}}
\caption{continued.}
\end{figure*}

\clearpage
\begin{figure*}
\addtocounter{figure}{-1}
\centerline{\resizebox{0.83\hsize}{!}{\includegraphics[angle=0]{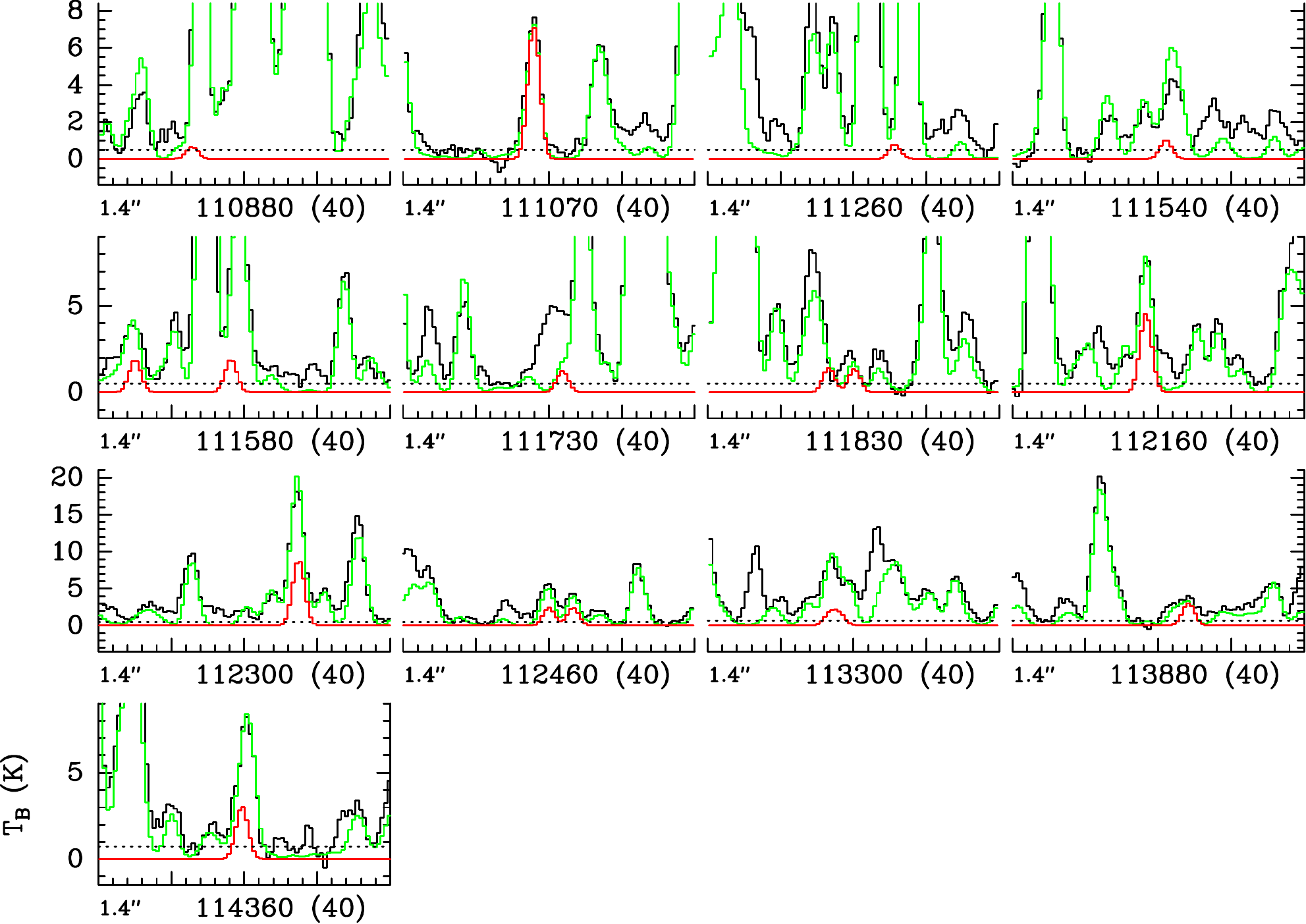}}}
\caption{continued.}
\end{figure*}

\clearpage
\begin{figure*}
\centerline{\resizebox{0.82\hsize}{!}{\includegraphics[angle=0]{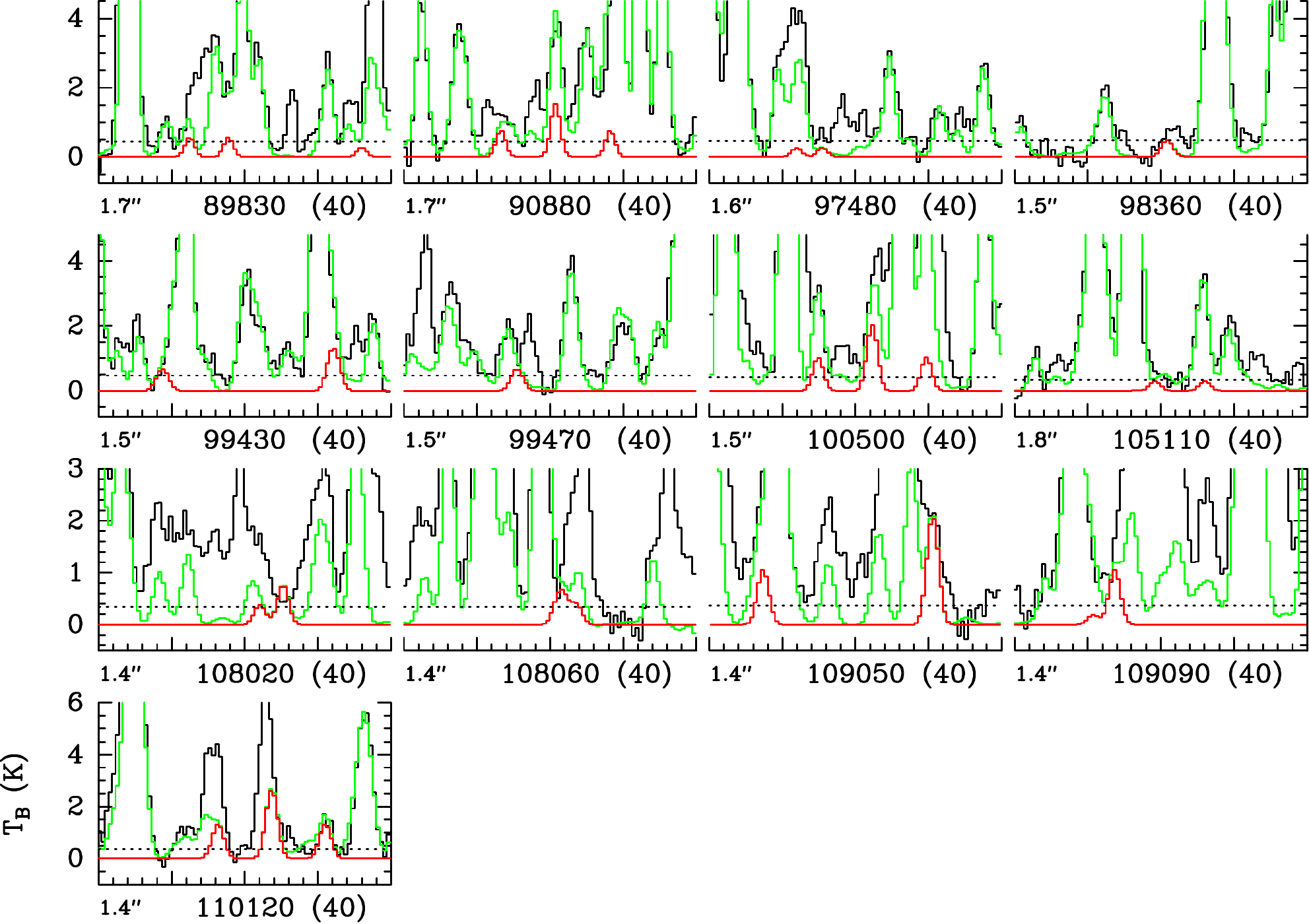}}}
\caption{Same as Fig.~\ref{f:spec_ch3coch3_ve0}, but for $^{13}$CH$_3$C(O)CH$_3$ 
$\varv = 0$.}
\label{f:spec_ch3coch3_13c1_ve0}
\end{figure*}

\clearpage
\begin{figure*}
\centerline{\resizebox{0.82\hsize}{!}{\includegraphics[angle=0]{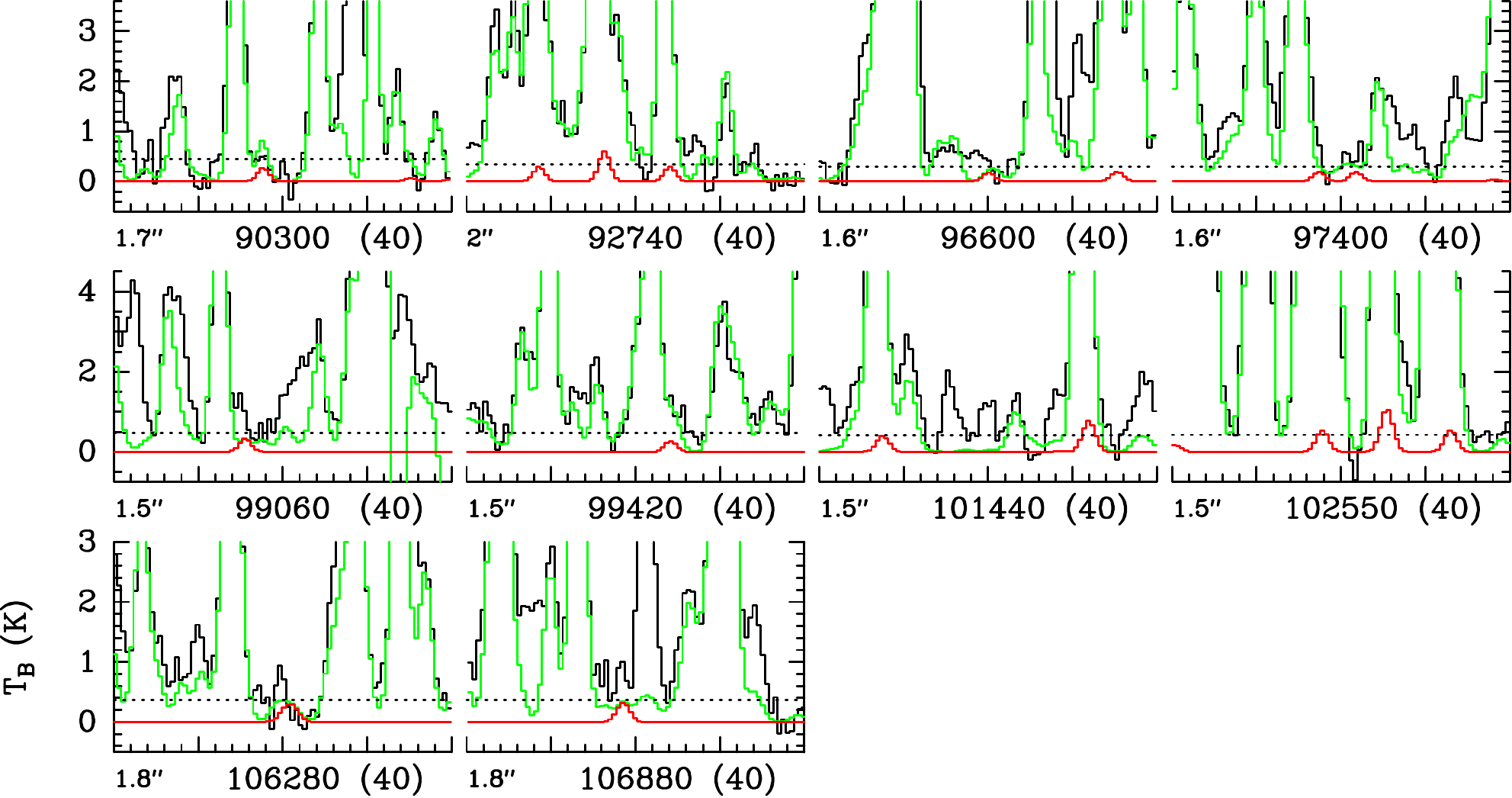}}}
\caption{Same as Fig.~\ref{f:spec_ch3coch3_ve0}, but for CH$_3^{13}$C(O)CH$_3$ 
$\varv = 0$.}
\label{f:spec_ch3coch3_13c2_ve0}
\end{figure*}

\end{appendix}


\end{document}